\documentclass[letterpaper, preprint,11pt]{AAS}
\usepackage{enumitem} 
\usepackage{mathtools}
\usepackage{gensymb}
\usepackage{float}
\usepackage{graphicx}
\DeclareGraphicsExtensions{.pdf,.png,.jpg} 
\usepackage{xcolor}
\usepackage{tabularx}
\usepackage[labelfont=bf,textfont=bf,font=footnotesize]{caption}
\usepackage{adjustbox}
\usepackage{amsmath}  
\usepackage{siunitx} 
\usepackage{amssymb}  
\usepackage{bm}
\usepackage{amsmath}
\usepackage{comment}
\usepackage{overcite}
\usepackage{footnpag}			      	
\usepackage{amssymb}
\usepackage{multicol}
\usepackage{multirow}
\usepackage{booktabs}
\usepackage{pstool}             
\usepackage[colorlinks=true, pdfstartview=FitV, linkcolor=black, citecolor= black, urlcolor= black]{hyperref}
\usepackage{subfig}
 
\usepackage[labelformat=simple]{caption}
\usepackage[letterpaper,margin=1in]{geometry}
\usepackage{booktabs}  
\usepackage{amsthm}
\usepackage{hyperref}
\hypersetup{
    colorlinks=true,
    urlcolor=black,
}

\PaperNumber{25-627}

\makeatletter
\newcommand{\ostar}{\mathbin{\mathpalette\make@circled*}}
\newcommand{\make@circled}[2]{%
  \ooalign{$\m@th#1\smallbigcirc{#1}$\cr\hidewidth$\m@th#1#2$\hidewidth\cr}%
}
\newcommand{\smallbigcirc}[1]{%
  \vcenter{\hbox{\scalebox{0.77778}{$\m@th#1\bigcirc$}}}%
}
\makeatother
\usepackage{url}
\begin{document}

\title{The Sustainability of the LEO Orbit Capacity via Risk-Driven Active Debris Removal}

\author{Yacob Medhin\thanks{PhD Student, Department of Aerospace Engineering, Iowa State University, IA 50011, USA. email: yacbin@iastate.edu} 
\ and Simone Servadio\thanks{Assistant Professor, Department of Aerospace Engineering, Iowa State University, IA 50011, USA. email: servadio@iastate.edu}
}

\maketitle
\begin{abstract}
The growing number of space debris in Low Earth Orbit (LEO) jeopardizes long-term orbital sustainability, requiring efficient risk assessment for active debris removal (ADR) missions. 
This study presents the development and validation of Filtered Modified MITRI (FMM), an enhanced risk index designed to improve the prioritization of high-criticality debris. 
Leveraging the MOCAT-MC simulation framework, we conducted a comprehensive performance evaluation and sensitivity analysis to probe the robustness of the FMM formulation. 
The results demonstrate that while the FMM provides superior identification of high-risk targets for annual removal campaigns, a nuanced performance trade-off exists between risk models depending on the operational removal cadence. 
The analysis also confirms that physically grounded mass terms are indispensable for practical risk assessment. 
By providing a validated open source tool and critical insights into the dynamics of risk, this research enhances our ability to select optimal ADR targets and ensure the long-term viability of LEO operations.
\end{abstract}
 
\section{Introduction}

Since the dawn of the space age in 1957, the number of objects in Earth's orbit has grown steadily, resulting in an increasingly crowded environment. 
Space surveillance networks currently track more than 30,000 pieces of debris over 10~cm in size within Earth's orbit, with more than half located in Low Earth Orbit (LEO) [\citen{ESA2024}]. 
This tally grows every year as launch rates accelerate; 2022 alone saw 2,409 new satellites launched, the most ever in a year [\citen{ESA2024}]. 
This surge is driven not only by commercial ventures and the deployment of mega satellite constellations in LEO but also by the increased launch cadence of small spacecraft over the past decade. 
For example, a single operator (SpaceX) has already deployed about 6,000 Internet satellites and plans to deploy tens of thousands more [\citen{britannica_starlink}]. 
Consequently, the exponential increase in orbital debris significantly increases collision risks for operational satellites, forcing them to carry out increasingly frequent and complex avoidance maneuvers [\citen{ESA2024}, \citen{bernat2020orbital}]. 
Collisions between orbiting objects can create additional debris, potentially triggering a cascading effect known as the Kessler syndrome, where each collision generates further fragments that endanger the entire orbital population and jeopardize the long-term viability of space activities [\citen{bernat2020orbital,catulo2023predicting,servadio2024threat}].

Numerous mission architectures for debris remediation and Active Debris Removal (ADR) have been proposed, ranging from ground-based lasers to non-contact techniques [\citen{mark2019review}]. 
The primary approach involves using a Robotic Servicing Vehicle (RSV) to capture and deorbit derelict objects [\citen{simha2025optimal}]. 
A crucial component of this strategy is determining which derelict objects to target to maximize future risk reduction.

To address this challenge, several risk indices have been developed based on static, or ``offline,'' parameters that describe an object's state at a single point in time. 
For example, the \emph{Criticality of the Spacecraft Index (CSI)} [\citen{rossi2015criticality}] assesses risk based on physical characteristics like mass, its orbital lifetime, and the local debris density. 
The \emph{Ranking Index ($R_N$)} [\citen{anselmo2015compliance}] builds on this by incorporating additional factors like debris flux, while the \emph{Environmental Consequences of Orbital Breakups (ECOB)} index [\citen{letizia2017extending}] integrates the probability and severity of fragmentation events over an object's lifetime. 
While these indices provide a valuable initial framework, their fundamental limitation is their static nature. 
They offer a single snapshot of risk and do not account for the constantly changing, stochastic characteristics of the orbital environment [\citen{servadio2024risk}].

To overcome these limitations, recent developments have focused on integrating dynamic, or ``online,'' components that continuously update an object's risk profile in real time. 
This approach considers the evolving orbital environment, including changes in collision probability, dynamic debris flux, and interactions with other satellites [\citen{servadio2024risk}]. 
For instance, following a breakup event, a sudden spike in local debris density can significantly elevate collision risks---an effect that static indices overlook. 
Advanced tools like MIT's Monte Carlo Orbital Capacity Assessment Tool (MOCAT-MC) employ this dynamic approach, using Monte Carlo simulations to model orbital perturbations, atmospheric drag, and stochastic collision events, thereby providing a more adaptive and predictive risk assessment [\citen{jang2025new}].

This work leverages the MOCAT-MC framework to introduce and validate a refined ranking model, the Filtered Modified MITRI (FMM). 
While MOCAT-MC provides the necessary dynamic modeling capability, this paper details the specific enhancements made to the risk index itself to more precisely estimate debris criticality. 
By validating these modifications against established scenarios and exploring their application to future LEO cases, this research provides a novel, open-source tool for prioritizing ADR targets to support long-term orbital sustainability.

\section{Methodology}
This section examines the approach for evaluating and prioritizing space debris for Active Debris Removal via MOCAT-MC. Current risk indices, such as CSI, offer a static, snapshot-based assessment of risk that fails to consider the dynamic nature of the debris environment. Conversely, MOCAT-MC incorporates dynamic risk factors, continuously updating real-time collision probabilities and orbital interactions to provide a more flexible rating. This section will initially highlight the simulation framework of MOCAT-MC, followed by a section on the computation of performance indices, comparing the static CSI with the MIT Risk Index (MITRI), which incorporates dynamic elements for risk evaluation. Finally, modified versions of MITRI are introduced, refining target selection through additional filtering techniques and enhanced weighting of key parameters to improve ADR decision-making and long-term space sustainability.
\subsection{Simulation Framework: MOCAT-MC}
\begin{figure}[htbp]
    \centering
    \includegraphics[width=\linewidth]{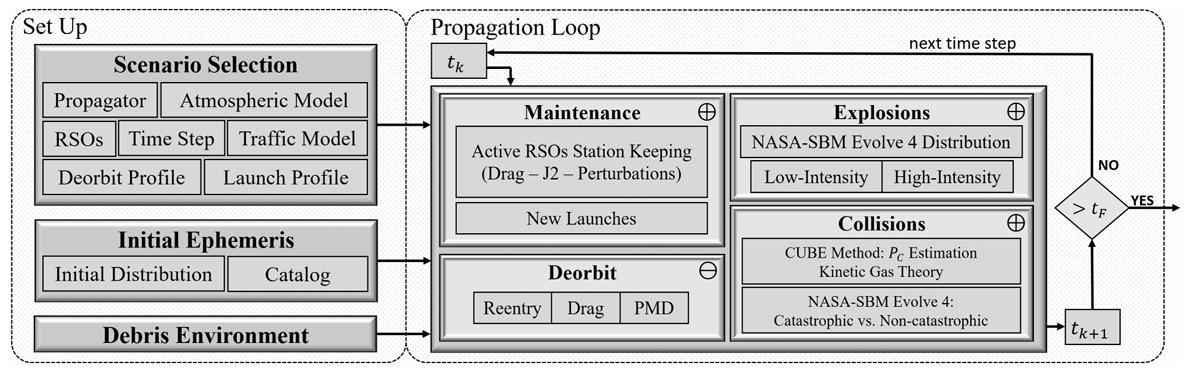}
    \caption{Schematic of the Monte Carlo tool MOCAT-MC [\citen{servadio2024risk}].}
    \label{fig:schematic}
\end{figure}

The MIT Monte Carlo Orbital Capacity Assessment Tool (MOCAT-MC) is a full-scale, three-dimensional debris evolutionary model designed to assess the Low Earth Orbit (LEO) population of Resident Space Objects (RSOs) [\citen{jang2025new}]. A schematic of the overall MOCAT-MC framework is shown in Figure \ref{fig:schematic}. Unlike earlier source-sink models of MOCAT [\citen{ashley2024parameters}, \citen{jang2024modeling}], which classify Active Space Objects (ASOs) into families and propagate their populations collectively, MOCAT-MC individually propagates each object and models its interactions at each time step. This approach allows for precise tracking of specific objects, accounting for individual characteristics such as mass, cross-sectional area, and orbital history. While computationally more expensive than its source-sink counterpart, MOCAT-MC enables a more detailed assessment by incorporating evasive maneuvers, orbit-keeping adjustments, and dynamic physical parameter changes over time. Developed in MATLAB as an open-source tool [\citen{MOCAT-MC}], MOCAT-MC provides the space-debris research community with a validated Monte Carlo-based framework to simulate the long-term evolution of the LEO environment, offering a higher-fidelity approach to debris mitigation analysis.
MOCAT-MC employs the Monte Carlo methodology, with each simulation run producing unique results based on stochastic variables and related probability density functions (PDFs). The major inputs to the model are: 
\begin{enumerate}[label={\alph*)}] 
    \item {Selection of a propagator}: MOCAT-MC can handle fast analytical, semi-analytical, and numerical propagators. The current approach employs a first-order analytical propagator that considers atmospheric drag effects (modeled using an exponential density profile) and Earth's J2 gravitational perturbation [\citen{martinusi2017first}].
    \item {Simulation parameters}: The duration of the simulation and time-step resolution (typically set to 5 days) influence computational efficiency and accuracy.
    \item {Initial debris population}: The model sets up objects with publicly available Two-Line Elements (TLEs) from sources such as ESA's DISCOS database\footnote{\url{https://discosweb.esoc.esa.int/}}  and the United States Space Force's Space-Track.org\footnote{\url{https://www.space-track.org/}}, resulting in a realistic initial debris environment [\citen{simha2025optimal}].
\end{enumerate}
During each time step, MOCAT-MC accounts for multiple stochastic space events, including:
\begin{enumerate}[label={\alph*)}] 
    \item {Station-keeping maneuvers and evasions}: Active spacecraft execute orbital adjustments and collision-avoidance maneuvers to prevent orbital decay and minimize collision risk. These adjustments are incorporated dynamically within the Monte Carlo simulation framework, allowing for realistic space traffic modeling [\citen{servadio2024risk}].
    \item {Collisions}: The collision probability is estimated using the NASA SMB Evolve 4 model  [\citen{johnson2001nasa}], where a uniform random number is drawn between 0 and 1. Suppose this number is lower than the computed collision probability. In that case, the model assumes the collision occurs, and the NASA Standard Breakup Model (SBM) is used to generate debris, assigning Two-Line Elements (TLEs) to each fragment.
    \item {Explosions}: Explosions are simulated as probabilistic failure events; objects undergo fragmentation based on historical failure rates. The number and characteristics of generated debris follow an empirical analysis model, ensuring realistic fragmentation outcomes [\citen{johnson2001nasa}].
    \item {New launches}: Future satellite deployments are modeled based on an expected annual launch rate, extrapolated from historical launch trends. The launch times, locations, and object properties are sampled from probabilistic distributions, reflecting uncertainty in future space traffic growth [\citen{servadio2024risk}].
    \item {Post-mission disposal}: Satellites are either successfully deorbited or fail disposal maneuvers at the end of their operational lifetime, with failure rates determined by empirical probability distributions derived from historical deorbit failures [\citen{liou2003new}]. Failed satellites remain as derelicts, contributing to the long-term debris population.
\end{enumerate}
MOCAT-MC uses Monte Carlo simulations and dynamic risk modeling to provide a complete framework for rating and selecting high-risk debris, which is an important step in minimizing the growing hazard of space debris and maintaining LEO sustainability.
\subsection{The MITRI Risk Index Formulation}
The performance indices (PI) presented in this article are built from previous work, where the criticality and danger of RSOs were estimated using empirical formulations. By analyzing previous approaches, we are able to develop a more inclusive PI that merges offline components, which are invariable among Monte Carlo simulations, such as mass, area, and persistence (residual lifetime), with online characteristics extracted at each time step from the MOCAT-MC, such as spatial density, probability of collisions, and number of debris generated.
Key indices such as Rossi's Criticality of Spacecraft Index (CSI), Anselmo and Pardini's Ranking Index (RN), and Letizia's ECOB were instrumental in shaping the design of the MIT Risk Index (MITRI), which we will use as our core PI. \(R_N\) [\citen{anselmo2015compliance}], a normalized dimensionless risk index, is a function of mass \(M\), altitude \(h\), and inclination \(i\). It is expressed as
\begin{equation}
\begin{aligned}
R_N = \frac{F}{F_0} \cdot \frac{l(h)}{l(h_0)} \cdot \left(\frac{M}{M_0}\right)^{1.75} \cdot \frac{CDCD50(h)}{CDCD50(h_0)} \cdot \frac{z(h,i)}{z(h,i=0)}
\end{aligned}
\end{equation}
The flux of debris capable of causing a catastrophic collision with the target object is represented by \(F\), while \(l_h\) denotes the lifetime function. The term \(CDCD50_h\)  refers to the concept of collisional debris cloud decay, representing the point at which \(50\%\) of the cataloged fragments (typically those of size \(d\geq10\)) in the Low Earth Orbit (LEO) have dissipated. The normalization factors are given by\( M_0=934\) \(kg\), \(h_0=800km\), and \(i_0=98.5^\circ\). The function \(z\) represents the flux capable of inducing catastrophic fragmentation at the orbital inclination of the tracked debris, normalized by the flux at zero inclination. This ratio accounts for objects in high-inclination orbits interacting with more objects than those in equatorial orbits. Thus, they are at increased risk of collisions.
\\CSI [\citen{rossi2015criticality}] by Rossi is expressed as
\begin{equation}
\begin{aligned}
CSI = \frac{M}{M_0} \cdot \frac{\rho_B(h)}{\rho_{B,0}} \cdot \frac{l(h)}{l(h_0)} \cdot g(i)
\end{aligned}
\end{equation}
where \( M_0 \) and \( h_0 \) represent the reference mass and altitude values with \( 10,\!000~\mathrm{kg} \) and \( 1,\!000~\mathrm{km} \), respectively. \( \rho_B(h) \) represents the spatial density, with \( \rho_{B,0} \) being its maximum value corresponding to the reference altitude of \( 770~\mathrm{km} \). The function \( g(i) \) adjusts for the influence of orbital inclination. Additionally, \( k \) is set to \( 0.6 \), reflecting that the typical debris flux in an almost equatorial orbit is approximately \( 60\% \) of the flux observed in a polar orbit.
\begin{equation}
\begin{aligned}
g(i) = \frac{1 + k\left(\frac{1 - \cos(i)}{2}\right)}{1 + k}
\end{aligned}
\end{equation}
\\ECOB [\citen{letizia2017extending}] by Letizia et al. is expressed as:
\begin{equation}
\begin{aligned}
ECOB = \int_{t_0}^{t_f} I \, dt \quad \text{where} \quad I = p_e \cdot e_e + p_c \cdot e_c
\end{aligned}
\end{equation}
ECOB quantifies the cumulative environmental impact of space objects. This metric, 
\( I \), combines the probabilities \( (p) \) of collisions \( (p_c) \) and explosions \( (p_e) \) with the corresponding environmental severity \( (e_e, e_c) \) of such events. Unlike previous static risk assessments, ECOB is computed throughout the operational life of the object (from \( t_0 \) to \( t_f \)) by integrating \( I \).\\
MITRI is a dynamic performance metric designed to assess the threat posed by space debris in the Low Earth Orbit (LEO) by integrating static and time-dependent factors. Developed as an advancement of existing indices (e.g., CSI, ECOB), MITRI quantifies debris criticality using six multiplicative terms:
\begin{equation}
\begin{aligned}
MITRI &= \left(\frac{M}{M_0}\right)^{1.75} \cdot \frac{\rho_B(h)g(i)}{\rho_{B,0}} \cdot \frac{L}{L_0} \cdot \mathbb{E}\left[\frac{R}{R_0}\right] \cdot \mathbb{E}\left[\frac{D}{D_0}\right] \cdot \mathbb{E}\left[\frac{P}{P_0}\right]
\end{aligned}
\end{equation}
In the above formulation, the terms R, D, and P represent the expected values averaged over time of the CUBE density, the yearly generation of debris, and the probability of collision, respectively, derived from the Monte Carlo simulations and evaluated for each spacecraft.
\subsubsection{Debris Mass}\leavevmode\newline
The mass term $\left(\frac{M}{M_0}\right)^{1.75}$ quantifies the threat posed by a debris object based on its mass ($M$). Heavier objects generate more fragments during collisions, with the exponent $1.75$ reflecting the combined probability ($1.0$) and consequence ($0.75$) factors from the NASA SBM, which empirically models the number of fragments generated as being proportional to the mass of the parent object raised to the power of 0.75 [\citen{mcknight2021identifying}]. $M_0$ is a normalization mass chosen based on mission parameters.
\subsubsection{Background Density}\leavevmode\newline
The term $\frac{\rho_B(h) \cdot g(i)}{\rho_{B,0}}$ evaluates the debris's orbital environment. $\rho_B(h)$ represents the spatial density of objects in the debris's altitude shell ($h$) within 50\,km altitude concentric shells in Low Earth Orbit (LEO), while $g(i)$ adjusts for inclination ($i$) effects, penalizing high-inclination orbits that intersect more satellites. $\rho_{B,0}$ normalizes the density using a reference value (e.g., median or maximum in the region). $\rho_B(h,t)$ evolves over time ($t$) due to launches, collisions, and decay, modeled via Monte Carlo simulations (e.g., MOCAT-MC), and is relevant to the RSO criticality [\citen{d2022analysis}].
\subsubsection{Residual Lifetime}\leavevmode\newline
The persistence term \(L/L_0\) estimates how long the debris remains in orbit. The numerical lifetime \(L\) is calculated by integrating orbital decay caused by atmospheric drag:
\begin{equation}
\begin{aligned}
\dot{a} = -BC \rho \sqrt{\mu a}
\end{aligned}
\end{equation}
where \(BC\) is the ballistic coefficient, \(\rho\) is the atmospheric density, and \(\mu\) is the gravitational parameter of the Earth. \(L_0\) normalizes the lifetime using median values.

\subsubsection{CUBE Density}\leavevmode\newline
The CUBE density term, formulated as\(\mathbb{E}\left[\frac{R}{R_0}\right]\), quantifies the expected value of localized orbital crowding, a key indicator of collision risk. In this expression, \(\mathbb{E}[...]\) is the expected value operator, which computes the average outcome of a random variable over many Monte Carlo simulations. The term \(R\) represents the time-averaged spatial density of objects within a discretized orbital volume (a CUBE), while \(R_0\) serves as a normalization factor.

The methodology assesses the localized collision risk by dividing the orbital space into volumetric cubes (e.g., 50 km\(^3\)) and applying the kinetic gas theory, where objects are treated analogously to gas molecules [\citen{lewis2019limitations}]. When integrated with NASA's EVOLVE \(4.0\) fragmentation model, this framework simulates collision outcomes (catastrophic or non-catastrophic) and subsequent debris generation. The implementation involves tracking object movements through these cubes and probabilistically evaluating collisions via Monte Carlo sampling [\citen{johnson2001nasa}]. This approach yields a computationally efficient (\(O(n)\)) forecast of localized crowding, which is critical for MITRI’s dynamic risk assessment.

\subsubsection{Probability of Collision}\leavevmode\newline
The collision probability term \(\mathbb{E}\left[\frac{P}{P_0}\right] \) in MITRI evaluates the likelihood of a debris object colliding with other RSOs over time. It employs a statistical approach based on the CUBE method, which discretizes orbital space into volumetric segments (e.g., 50 km\(^3\) cubes) and applies kinetic gas theory to estimate collision risks.

The overall probability of collision of an RSO is evaluated according to the following procedure:
\begin{enumerate}[label={\alph*)}]
    \item{Pairwise Collision Probability (\(p_j\))}: \\
    The probability of collision between a tracked object \(i\) and another object \(j\) within the same cube is given by:
    \begin{equation}
    \begin{aligned}
    P_{i,j} = s_i s_j V \sigma dU
    \end{aligned}
    \end{equation}
    where \(s_i\) and \(s_j\) denote the spatial densities of objects \(i\) and \(j\) within the cube, \(V\) is the relative velocity between the objects, \(\sigma\) is the cross-sectional area of collisions, and \(dU\) is the volume of the cube.
    \item {Total Collision Probability (\(P_C\))}\\[1mm]
    The overall probability that the tracked object collides with any object in its cube is given by: \begin{equation}
    P_C = 1 - \prod_{j=1}^{N} (1 - p_j)
    \end{equation}
    This assumes the independence between collision events (a conservative estimate) and indicates the probability of the RSO to collide, in general, with any other RSO at a specific time. 
    \item {Time-Averaged Probability (\(P\))}\\[1mm]
    MITRI calculates the expected value \(\mathbb{E}[P]\) through a multi-step process to account for the time-varying nature of collision likelihood. First, Monte Carlo simulations are executed using the MOCAT-MC framework to generate probabilistic collision data. These results are then fitted to an exponential distribution [\citen{hernandez2018comparison}], formally expressed as \begin{equation}
    P_C(x) = \lambda \exp(-\lambda x) \quad \forall x > 0
    \end{equation}
    where \(\lambda\) is the distribution rate parameter. The mean probability \(P\), derived from the fitted distribution, is subsequently used for normalization and is defined as the reciprocal of the rate parameter:
    \begin{equation}
        P = \frac{1}{\lambda}.
    \end{equation}
    This approach ensures that the time-averaged collision probability is consistently scaled and interpretable across varying operational scenarios.
\end{enumerate}
This method provides a computationally efficient (\(O(n)\)) approach to assess long-term collision risks while addressing the evolving space environment. This dynamic approach represents a significant improvement over static risk metrics.
\subsubsection{Yearly Generated Debris}\leavevmode\newline
The yearly generated debris term \(\mathbb{E}\left[\frac{D}{D_0}\right]\) quantifies the amount of new debris a space object is expected to produce annually through collisions or explosions, providing crucial insight into long-term environmental impact. The calculation employs NASA’s Standard Breakup Model (EVOLVE 4.0) [\citen{johnson2001nasa}] to determine fragment generation, with catastrophic collisions producing debris according to:
\begin{equation}
N_C = 0.1\, L_C^{-1/2} \, (M_i + M_j)^{0.75} 
\end{equation}
where \(L_C\) represents the characteristic length (km), \(N_C\) is the number of generated fragments, and \(M_i, M_j\) (kg) are the masses of the colliding objects. For non-catastrophic collisions, the model uses:
\begin{equation}
N_{NC} = 0.1\, L_C^{-1/2} \, \left(M_p \cdot v_{imp}^2\right)^{0.75} 
\end{equation}
accounting for the smaller object's mass (\(M_p\)) and impact velocity (\(v_{imp}\)) in km/s. Collisions are evaluated using the CUBE method and are categorized as damaging or catastrophic when the impact energy exceeds 40 J/g [\citen{liou2003new}, \citen{liou2006collision}]. The annual debris rate \(D\) is then derived by dividing the total simulated debris by the duration of the simulation (in years), allowing consistent comparisons across different mission timelines. This term is implemented through Monte Carlo simulations, probabilistically evaluating collision outcomes and potential explosion scenarios, with results normalized against a median value \(D_0\) to facilitate relative risk assessment.

\subsection{Enhanced Dynamic Risk Index}
Building on the original formulation in which MITRI uses three offline (static) and three online (dynamic) terms to measure the risk of space debris, we propose a fundamental enhancement to the way dynamic factors are calculated. This involves shifting from the post-simulation analysis of the original published MITRI [\citen{servadio2024risk}], which derived expected values by fitting a complete simulation data set to statistical models, to a concurrent analysis approach. In this new framework, the dynamic terms are calculated iteratively. The expected values for CUBE density (\(R\)), yearly generated debris (\(D\)), and the probability of collision (\(P\)) are updated at each time step using a simplified smoothed average:
\begin{equation}
E_{\text{n}} = \frac{E_{\text{n-1}} + x_{\text{n}}}{2}
\label{eq:smoothed_average}
\end{equation}
where \(E_{n-1}\) is the accumulated expected value from all prior time steps and \(x_n\) is the current value of the term observed at time step \(n\).

This concurrent framework is a significant methodological improvement and serves as the new baseline for the risk indices evaluated in this article, including the updated MITRI. Building upon this, we then introduce the Filtered Modified MITRI (FMM), an enhanced index that incorporates two further key modifications: the use of `fictitious collisions' to estimate debris from potential, high-probability fragmentation events. These ``what-if" scenarios help reveal the true criticality of nonactive satellites by capturing the additional debris that could be generated under realistic conditions and refining the yearly debris generation term in MITRI. Simultaneously, we replace the static background density assumption with a time-varying approach, recalculated several times a year in concentric orbital shells to reflect the constantly evolving LEO population, whether from collisions, explosions, new launches, or post-mission disposals. By continuously recalculating the background density of the LEO region rather than depending on an initial fixed value, we hope to obtain a more accurate view of the risk levels at any given time, improving MITRI's prediction power and capacity to identify and prioritize high-risk debris targets in ADR missions.

\subsubsection{ Fictitious Debris Generation and Filtering}\leavevmode\newline
The MITRI framework computed the yearly generated debris term, \( D \), based on stochastic events. In that model, the \(D\) term was only incremented after a collision was considered to have occurred, which happened when a pair's collision probability (\( P_{\text{coll}} \)) exceeded a randomly sampled threshold (\( \texttt{rand\_P} \)) during a timestep. This event-based approach has a significant limitation: it ignores the latent danger posed by objects in persistently crowded regions that have a high probability of collision but have not yet been part of a stochastically triggered event. To resolve this limitation, two sequential refinements were integrated:
\begin{enumerate}
  \item Fictitious Collision Modeling: This enhancement redefines the calculation of the Yearly Generated Debris term (\(D\)) by shifting from the stochastic, event-based approach to a proactive, potential-based one. At every time step, all potential conjunctions (close-approach pairs) are identified using the CUBE method, which discretizes LEO into volumetric segments. Then, for each identified pair, the potential number of fragments is calculated using the NASA SBM, regardless of whether a collision physically occurs. Mathematically, the \(D\) term is therefore calculated as the sum of all potential debris generated from all identified conjunctions over the simulation time. This ensures that objects in high-risk situations contribute to the risk index at every time step, reflecting their latent threat.
  \item {Debris Filtering}: Excludes fragments with mass \( \leq 10 \, \text{kg} \) or characteristic length \( \leq 10 \, \text{cm} \), aligning with NASA’s definition of ``trackable debris.'' This reduces computational noise from insignificant fragments.
\end{enumerate}

Collectively, these two refinements transform the Yearly Generated Debris (\(D\)) term into a more robust and forward-looking metric. The fictitious collision modeling provides a proactive assessment by capturing the latent threat from all high-probability conjunctions, while the subsequent debris filtering refines this calculation by reducing computational noise from insignificant fragments. This architectural change fundamentally shifts the risk framework from being reactive to specific, randomly triggered events to being proactive about cumulative, potential risk. 

\subsubsection{Dynamic Background Density}\leavevmode\newline
Conventional risk indices, such as the original MITRI formulation, rely on static background density estimates that overlook the dynamic change in the LEO environment. Hence, a time-dependent strategy is required to preserve accuracy in risk assessment. To overcome this constraint, we present a time-varying background density framework regenerated at regular intervals (1, 3, and 6 months) during a simulation duration of 200 years. This improvement guarantees that MITRI will actively adjust to population fluctuations, enabling it to prioritize high-risk ADR goals.\\
The methodology shifts from static to dynamic background density estimate using the previous architecture from MOCAT-MC implementations, where the LEO region is divided into concentric spherical shells at 50 km altitude increments. The spatial density $\rho_B(h, t)$ for each shell is recalculated at configurable intervals ($\Delta t = 1, 3, 6$ months) over a 200-year simulation using [\citen{d2022analysis}]:
\begin{equation}
\rho_B(h, t) = \frac{N(h, t)}{V(h)},
\end{equation}
where $N(h, t)$ is the instantaneous object count at altitude $h$ and time $t$, and $V(h)$ is the shell volume. Stochastic events influencing $N(h, t)$, including collisions (modeled by NASA’s Standard Breakup Model), explosions, new launches, and post-mission disposals (successes/failures based on empirical rates)—are aggregated at each $\Delta t$.\\ The implementation integrates this recalculation into MOCAT—changes are evaluated at every interval; that is, $N(h, t)$ is updated and $\rho_B(h, t)$ is refreshed. This adaptive approach ensures that MITRI remains in sync with the evolving LEO environment.
\subsection{ Experimental Design and Results for Performance and Sensitivity Analysis}
This section details the experimental framework designed to rigorously evaluate the proposed Filtered Modified MITRI (FMM) index. Leveraging the MOCAT-MC simulation environment, we have structured a series of numerical experiments with a dual objective. The first objective is to establish the baseline performance of FMM by comparing its long-term impact on the LEO population against existing indices and a no-debris removal scenario. The second, more extensive objective is to conduct a comprehensive sensitivity analysis to probe the robustness of the FMM formulation.
This analysis systematically varies key model assumptions, including the removal cadence, the formulation of the mass term, the application of probabilistic epsilon scaling, and debris filtering thresholds, to quantify the contribution of each component to the index's overall effectiveness. By exploring these variations, we can ascertain the stability of our proposed model and identify the parameters most critical to its predictive power. To ensure the statistical validity of our conclusions, all scenarios were simulated over a 200-year horizon (unless otherwise specified), with final results representing the averaged outcome of multiple independent Monte Carlo runs. The following subsections outline the specific methodology for each experiment.

\subsubsection{Validation of Risk-Based ADR vs. Random Removal}\leavevmode\newline
Before evaluating the enhancements of the FMM index, it is essential first to validate the foundational premise of risk-driven Active Debris Removal. To this end, an initial simulation was conducted to compare the long-term effectiveness of a targeted removal policy, guided by MITRI, against a non-targeted, uniform random removal policy (RANDk).
This experiment simulates the evolution of the LEO population under several parallel scenarios: a ``no-ADR" baseline and scenarios where k = 1, 3, and 5 objects are removed annually according to either the MITRI rankings or the RANDk selection. 
\vskip -1em
\begin{figure}[H]
    \centering
    \subfloat[Random removal]{%
        \includegraphics[width=0.49\textwidth]{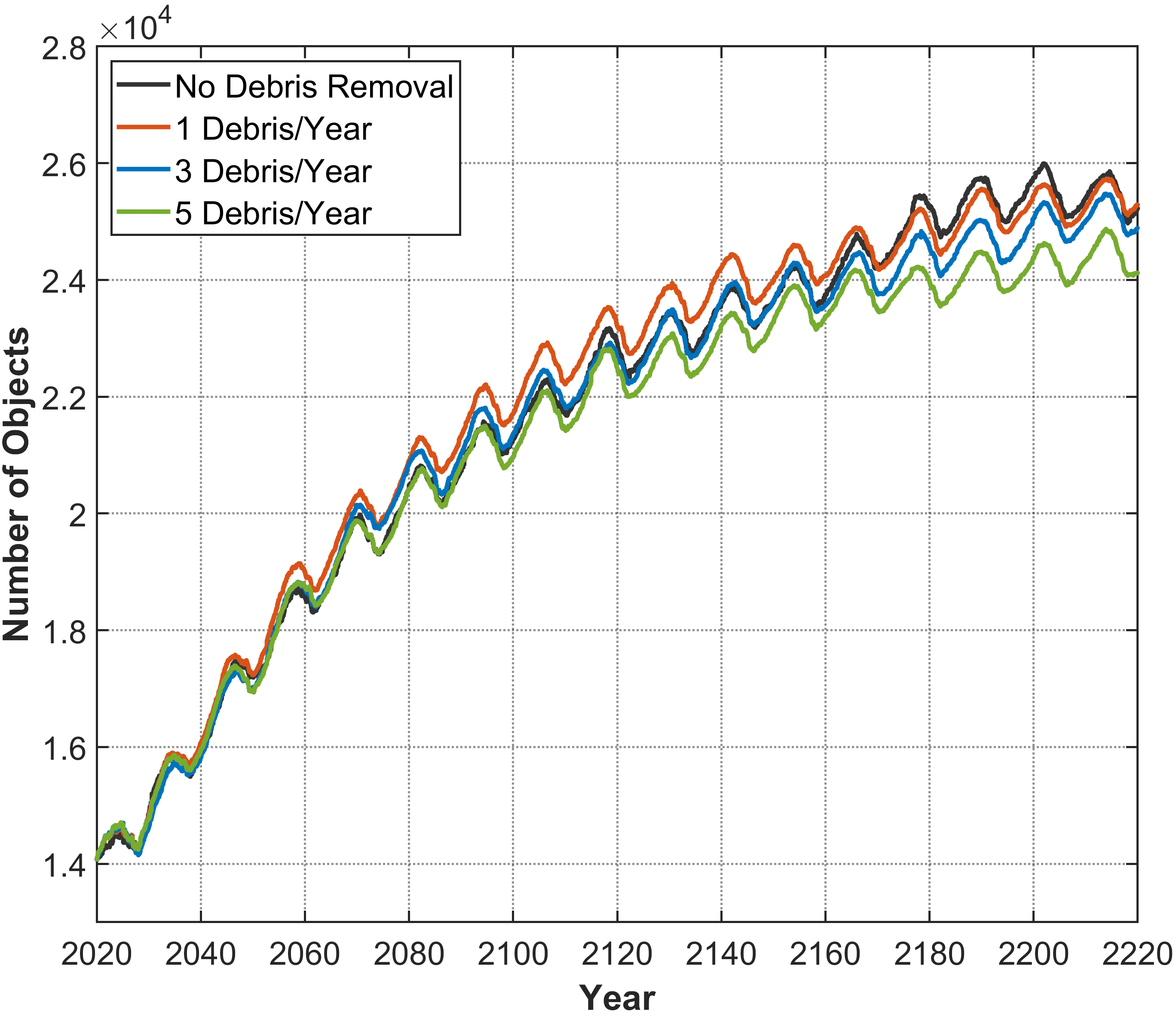}
        \label{fig:random}
    }
    \hfill
    \subfloat[Removing satellites with the highest MITRI value]{%
        \includegraphics[width=0.49\textwidth]{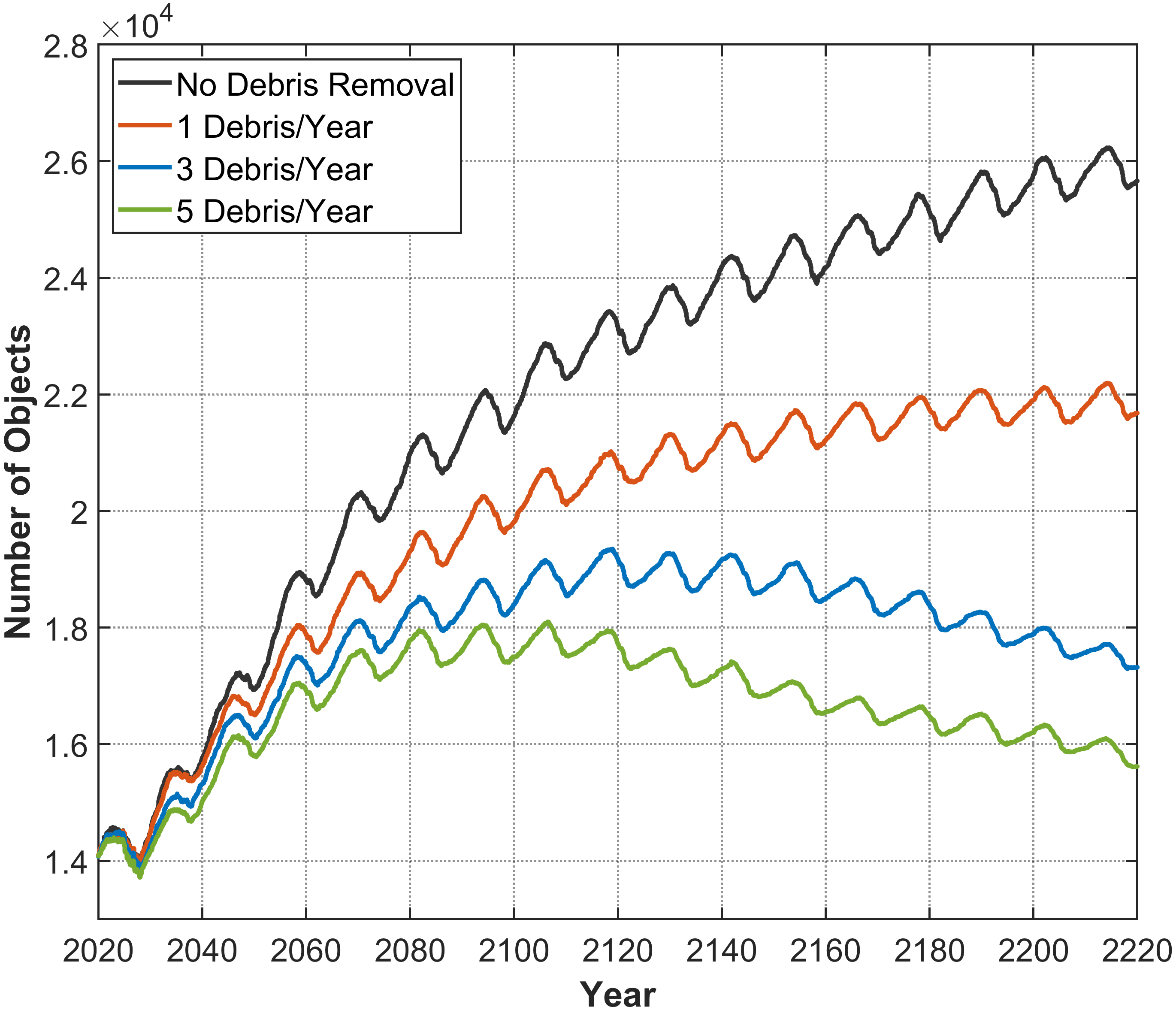}
        \label{fig:mitri}
    }
    \caption{Evolution of the LEO Population under Different Removal Policies.}
    \label{fig:mitri_vs_rand}
\end{figure}
\vskip -1em
The results of this foundational experiment, shown in Figure~\ref{fig:mitri_vs_rand}, confirm the premise of this entire research effort. As shown in Figure~\ref{fig:random}, the random removal of 1, 3, or 5 objects per year provides only a marginal deviation from the baseline scenario where no debris is removed. In contrast, Figure~\ref{fig:mitri} shows that a targeted removal policy guided by a risk index yields a substantial reduction in population growth. In particular, removing only one high-risk object per year is more effective in stabilizing the environment than randomly removing five objects per year. This result empirically demonstrates the fundamental principle that a targeted, risk-based approach provides a significant and quantifiable advantage over indiscriminate removal, establishing the critical importance of optimizing the risk index itself.

\subsubsection{Comparative Analysis of the FMM Index}\leavevmode\newline
Having established that a risk-based strategy is fundamentally superior to random removal, the critical next step is determining which risk index is the most precise. This experiment is designed to directly compare the ranking accuracy of our proposed FMM index against its predecessors, MITRI and the static CSI. We aim to answer the central question: ``How effectively does each index identify the specific objects that are statistically most likely to be involved in future collisions?"

\begin{figure}[H]
        \centering
        \subfloat[Linear‑scale plot (y=x reference line)]{%
            \includegraphics[width=0.49\textwidth]{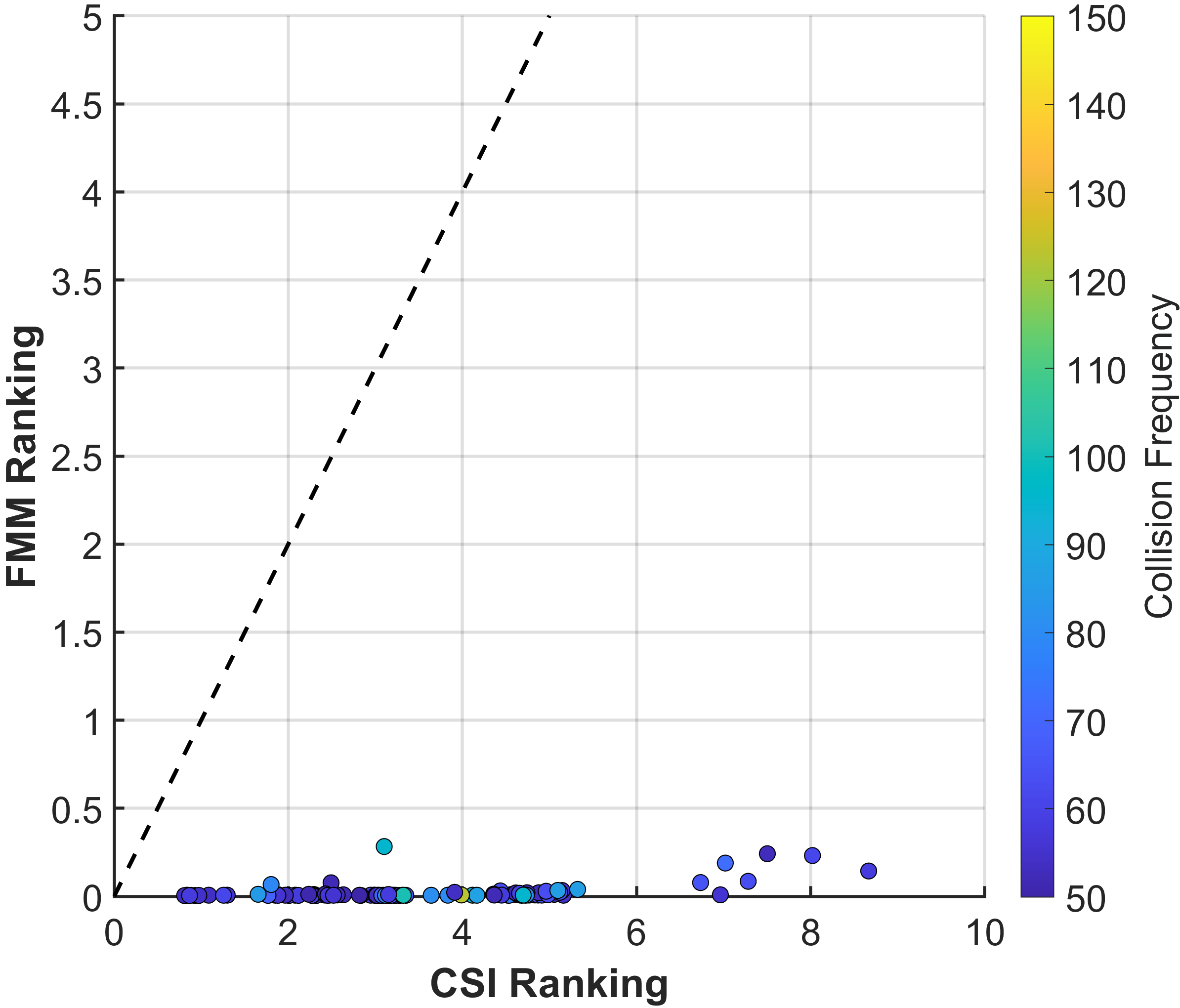}
            \label{fig:linear}
        }
        \hfill
        \subfloat[Semi‑log plot with logarithmic y‑axis (equality curve)]{%
            \includegraphics[width=0.49\textwidth]{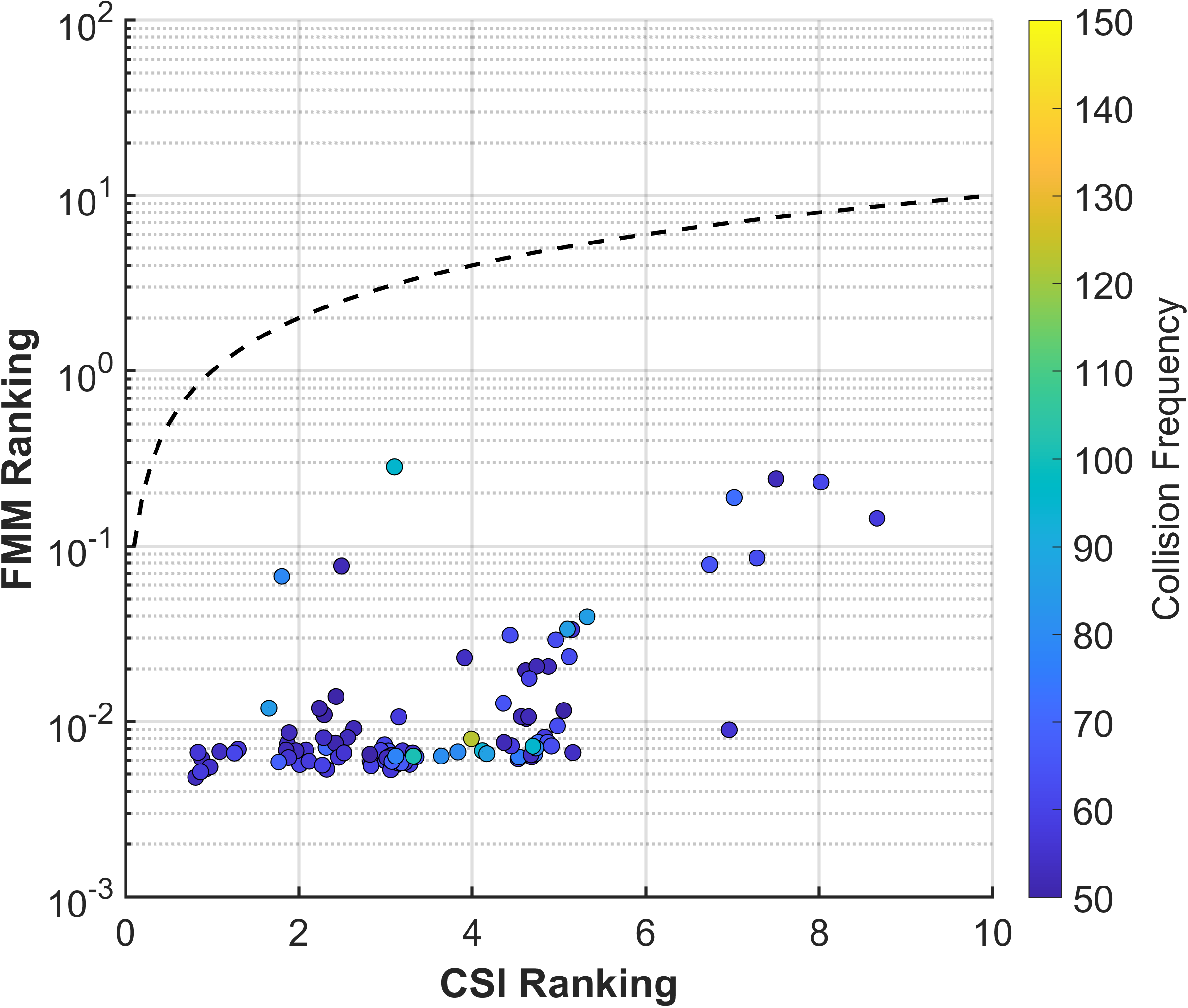}
            \label{fig:log_scale}
        }
        \caption{Evolution of the LEO Population under Different Removal Policies.}
        \label{fig:comparison1}
    \end{figure}

\begin{figure}[H]
        \centering
        \subfloat[Linear-scale distribution]{%
            \includegraphics[width=0.49\textwidth]{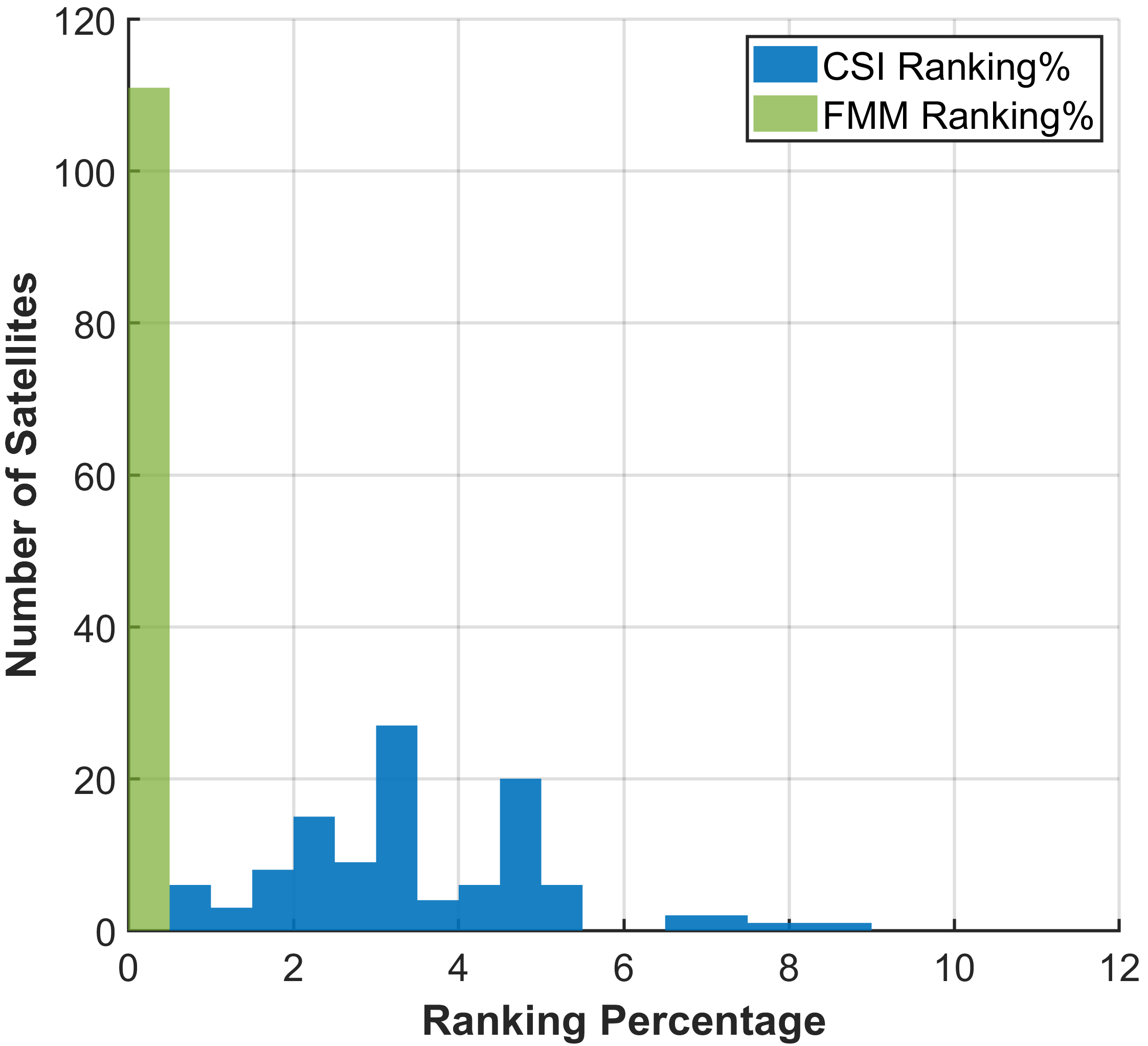}
            \label{fig:standard_distribution}
        }
        \hfill
        \subfloat[Log-scale distribution]{%
            \includegraphics[width=0.49\textwidth]{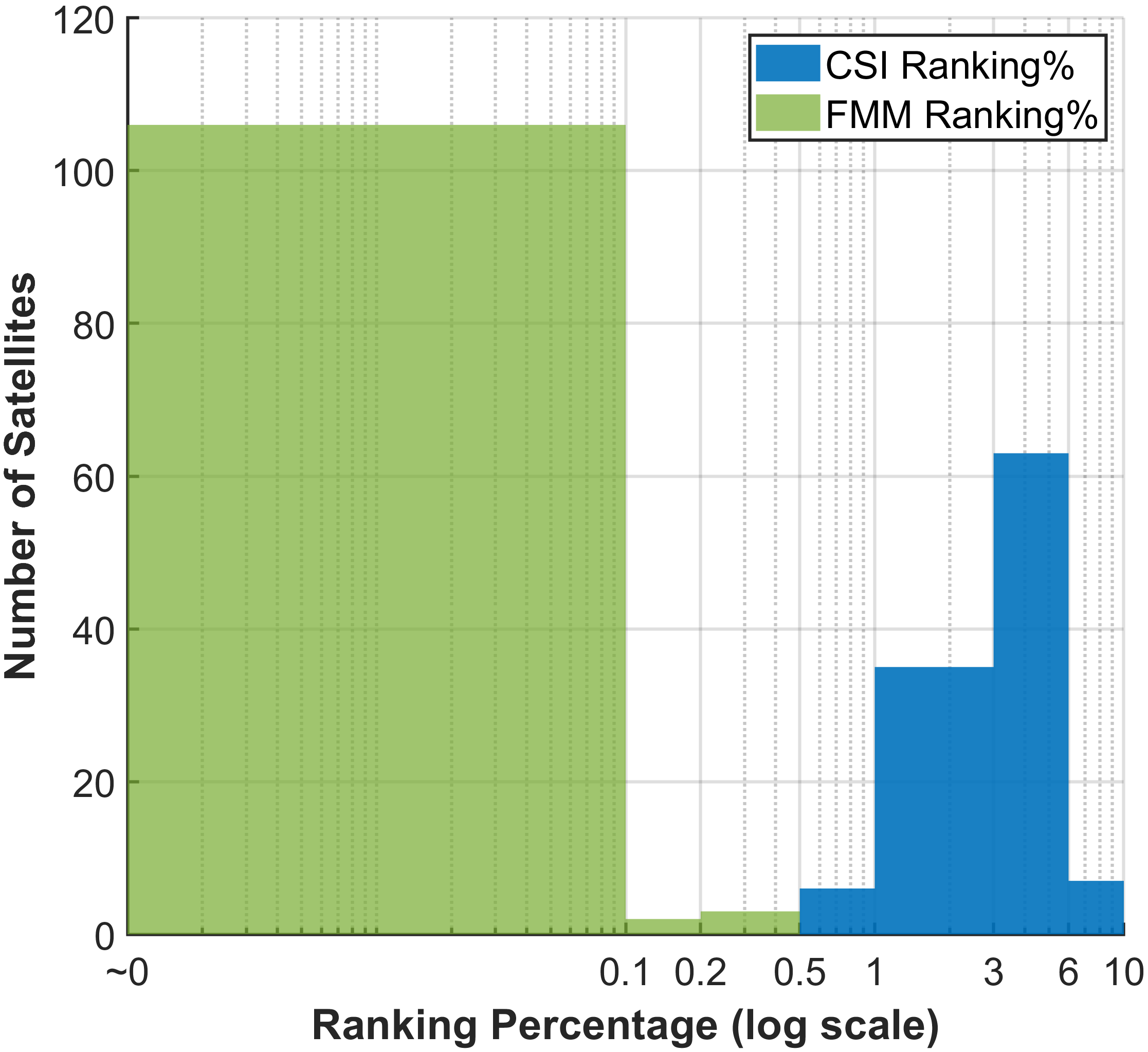}
            \label{fig:log_scale_distribution}
        }
        \caption{Evolution of the LEO Population under Different Removal Policies.}
        \label{fig:comparison2}
    \end{figure}

To objectively measure this, a ``ground truth" of verifiably dangerous objects was first established. This was achieved by conducting a comprehensive Monte Carlo campaign, running over 1000 unique seeds of the 200-year MOCAT-MC simulation. From this extensive dataset, we analyzed the output to identify any satellite that experienced 50 or more collision events in any given simulation run. This large-scale approach ensures that our resulting ``ground truth" cohort is not an artifact of a single stochastic timeline but instead represents a statistically robust set of objects with a repeatedly demonstrated, high potential for future collisions. With this ground truth cohort identified, we then performed a comparative analysis to evaluate how accurately each index, FMM, MITRI, and CSI, prioritized these proven threats. This evaluation was conducted in two parts:
\begin{enumerate}
   \item  {FMM vs. CSI Ranking Philosophy (Visualized in Figures~\ref{fig:comparison1} and ~\ref{fig:comparison2})}: To highlight the fundamental difference between a dynamic and a static approach, we directly plotted the ranking percentages assigned by FMM versus those assigned by CSI for each object in the high-risk cohort. The results, visualized in Figures~\ref{fig:comparison1} and~\ref{fig:comparison2}, confirm the superiority of the dynamic philosophy. Figure~\ref{fig:comparison1} provides a direct comparison for the high-risk satellite cohort. Data points consistently fall below the y=x equality line and the log y curve, indicating that FMM assigns more critical (i.e., lower) rankings to these proven threats than static CSI.
   
   This trend is further detailed in the ranking distribution shown in Figure~\ref{fig:comparison2}. FMM group high-risk objects into a tight distribution below a 0.5\% ranking, whereas the CSI rankings are more broadly distributed. This visual evidence establishes the superiority of a dynamic risk assessment framework, justifying our subsequent comparison between the proposed FMM index and its dynamic predecessor, MITRI, to quantify the benefits of our enhancements.
    \item {Core Performance Analysis: FMM vs. MITRI (quantified in Table~\ref{tab:identification_rates})}: The core experiment of this study directly compares the performance of the proposed FMM index against its predecessor, MITRI. The comparison is quantified using the ``High-Risk Identification Rate''—the percentage of ``ground truth'' objects that each index successfully ranks as highly critical (within the top 0.5\%). This analysis was conducted across multiple scenarios where the background density term for both indices was updated at different intervals (once, every 6 months, every 3 months, and every month) to assess how effectively each index leverages timely environmental data. This detailed comparison, presented in Table ~\ref{tab:identification_rates}, serves a dual purpose; firstly, it quantifies the superiority of FMM over MITRI, and then it assesses how the performance of both dynamic indices is affected by the frequency of their environmental updates.
    \begin{table}[htbp]
        \centering
        \caption{Comparison of FMM and MITRI High-Risk Identification Rates under varying background density update intervals. The values represent the percentage of high-collision satellites ($\ge 50$collisions) ranked within the top 0.5\%.}
        \label{tab:identification_rates}
        \begin{tabular}{l c c}
            \toprule
            \textbf{Update Interval} & \textbf{MITRI} & \textbf{FMM} \\
            \midrule
            1 time (Static) & 89.90\% & 98.15\% \\
            6 months        & 84.04\% & 98.13\% \\
            3 months        & 83.33\% & 99.99\% \\
            1 month         & 87.37\% & 98.08\% \\
            \bottomrule
        \end{tabular}
    \end{table}

    The results clearly show that FMM consistently outperforms MITRI across all tested cadences, achieving near-perfect identification rates (98.08\%–100\%). Although MITRI's performance is strong in a static configuration (89.90\%), its effectiveness degrades with more frequent updates—a counterintuitive result for a dynamic index. In contrast, FMM’s performance improves with more frequent data, peaking at a 100\% identification rate with a 3-month update interval. This demonstrates FMM’s superior ability to leverage timely data to more accurately identify high-risk targets.
\end{enumerate}
    \vskip -1em
    \vskip -1em
    \vskip -1em
    \vskip -1em
\subsubsection{Epsilon Scaling Implementation}\leavevmode\newline
In the standard FMM formulation, the term yearly generated debris ($D$) is calculated using a ``fictitious collision'' model. 
This approach assumes that every potential collision pair identified by the CUBE method at a given timestep contributes its full unweighted debris count to the $D$ term. 
Although this method effectively highlights high-risk areas, it treats a conjunction with a low collision probability ($P_{\text{coll}}$) as having the same immediate impact on the term $D$ as a conjunction with a very high $P_{\text{coll}}$.

To investigate whether a more nuanced, probability-weighted approach could improve target prioritization, we implemented and tested two distinct \emph{epsilon scaling} methodologies, designated $\varepsilon_1$ and $\varepsilon_2$. 
These techniques refine the fictitious collision model by scaling the potential debris contribution of each pair by a factor $\varepsilon$, which is a function of the pair's specific collision probability ($P_{\text{coll}}$) and the stochastic outcome of the simulation (\texttt{rand\_P}). The two models tested were:

\begin{enumerate}
    \item Non-linear Sigmoid Model, $\epsilon_1$: This model was chosen to represent a smooth, non-linear transition of risk, as shown in Equation~\eqref{eq:epsilon1}. The rationale for its form is as follows:
    \begin{itemize}
        \item The sigmoid function is a standard mathematical tool used in statistical modeling and machine learning to represent a probabilistic transition between two states, in this case, from near-zero risk contribution to full contribution [\citen{bishop2006pattern}]. Its S-shape ensures that $\varepsilon_1$ increases smoothly as the collision probability $P_{\text{coll}}$ approaches the random number \texttt{rand\_P}, which is a more physically plausible representation of an increasing risk than a simple step function.
        \item The $\sqrt{P_{\text{coll}}}$ baseline term acts as a sub-linear dampening factor. This ensures that objects with very high collision probabilities do not completely dominate the risk index in a single timestep, allowing for a more balanced assessment of moderate-risk versus high-risk pairs.
    \end{itemize}
    \begin{equation}\label{eq:epsilon1}
         \varepsilon_1 = \frac{1}{1 + e^{10(\texttt{rand\_P} - P_{\text{coll}})}} + k \sqrt{P_{\text{coll}}}, \quad \text{where} \quad k = \frac{\varepsilon_{\text{max}}}{\sqrt{P_{\text{max}}}}
    \end{equation}
\vskip -1em
\begin{figure}[H]
    \centering
    \includegraphics[width=0.6\textwidth]{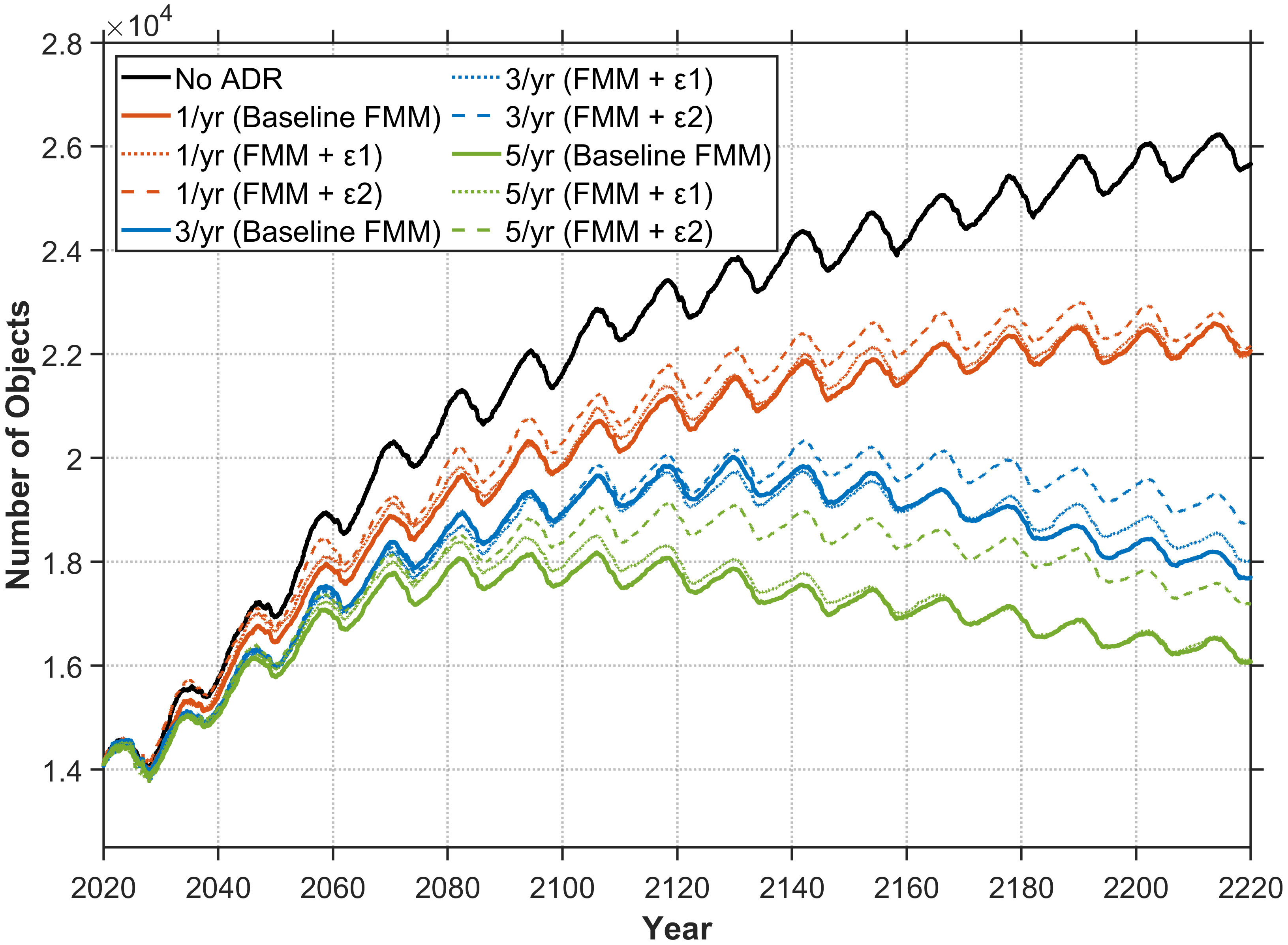} 
    \caption{Impact of Epsilon Scaling on LEO Population Evolution.}
    \label{fig:epsilon_scaling_results}
\end{figure}
\vskip -1em
    \item Linear Model, $\epsilon_2$: This model was chosen for its simplicity and interpretability, representing a first-order linear approximation of the risk scaling, as shown in Equation~\eqref{eq:epsilon2}. 
    This method provides a direct relationship where the assigned risk fraction scales linearly as $P_{\text{coll}}$ approaches the stochastic threshold \texttt{rand\_P}. 
    This simpler model serves as a crucial baseline to determine if the additional complexity of the non-linear sigmoid model ($\varepsilon_1$) provides any tangible benefit.
   \begin{equation}\label{eq:epsilon2}
        \varepsilon_2 = 1 - \frac{|\texttt{rand\_P} - P_{\text{coll}}|}{\texttt{rand\_P}}
    \end{equation}
\end{enumerate}
The underlying hypothesis for this experiment is that by weighting the fictitious debris contribution by its likelihood, the FMM index would more accurately reflect the true, probabilistic nature of the risk, leading to improved prioritization.

The results of this analysis are presented in Figure~\ref{fig:epsilon_scaling_results}. Contrary to the initial hypothesis, the key finding is that the implementation of epsilon scaling, in both forms, consistently degrades the performance of the ADR strategy compared to the baseline FMM. For any given removal rate $k$, both epsilon scenarios resulted in a higher final object count than their corresponding baseline run. Furthermore, the relative performance between the two epsilon models is consistent. In the removal cases $k=1$ and $k=3$, the linear scaling of $\varepsilon_2$ resulted in a marginally worse outcome than the hybrid sigmoid/baseline model of $\varepsilon_1$; however, this underperformance becomes more pronounced for the scenario $k=5$, where a larger divergence between the two models is evident.
\subsubsection{Debris Removal Cadence Analysis}\leavevmode\newline
This analysis investigates the strategic implications of removal timing on the long-term evolution of the debris environment. It addresses the operational question of whether it is more effective to conduct frequent, small-scale ADR campaigns or larger, less frequent ones. To quantify this, the baseline annual removal policy was compared against two extended-cadence scenarios, with removals of the top-ranked $n$ objects (where $k$ was set to $1, 3$, and $5$) occurring every $5$ and $10$ years. This experiment was performed for removals guided by both the MITRI and the enhanced FMM index.
\vskip -1em
\begin{figure}[H]
    \centering
    \subfloat[Annual vs. 5-Year Cadence]{%
        \includegraphics[width=0.49\textwidth]{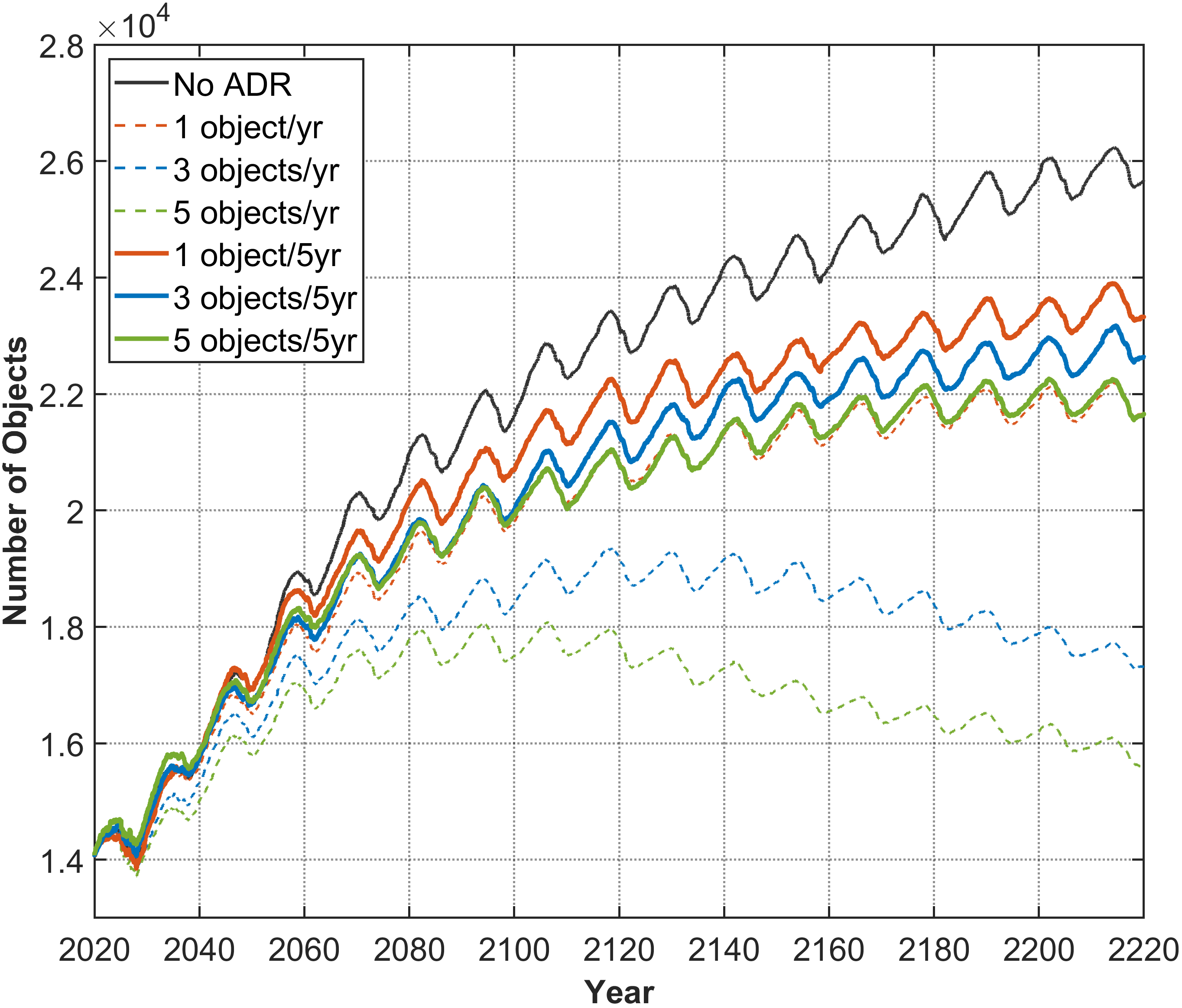}
        \label{fig:mitri_vs_5yr}
    }
    \hfill
    \subfloat[Annual vs. 10-Year Cadence]{%
        \includegraphics[width=0.49\textwidth]{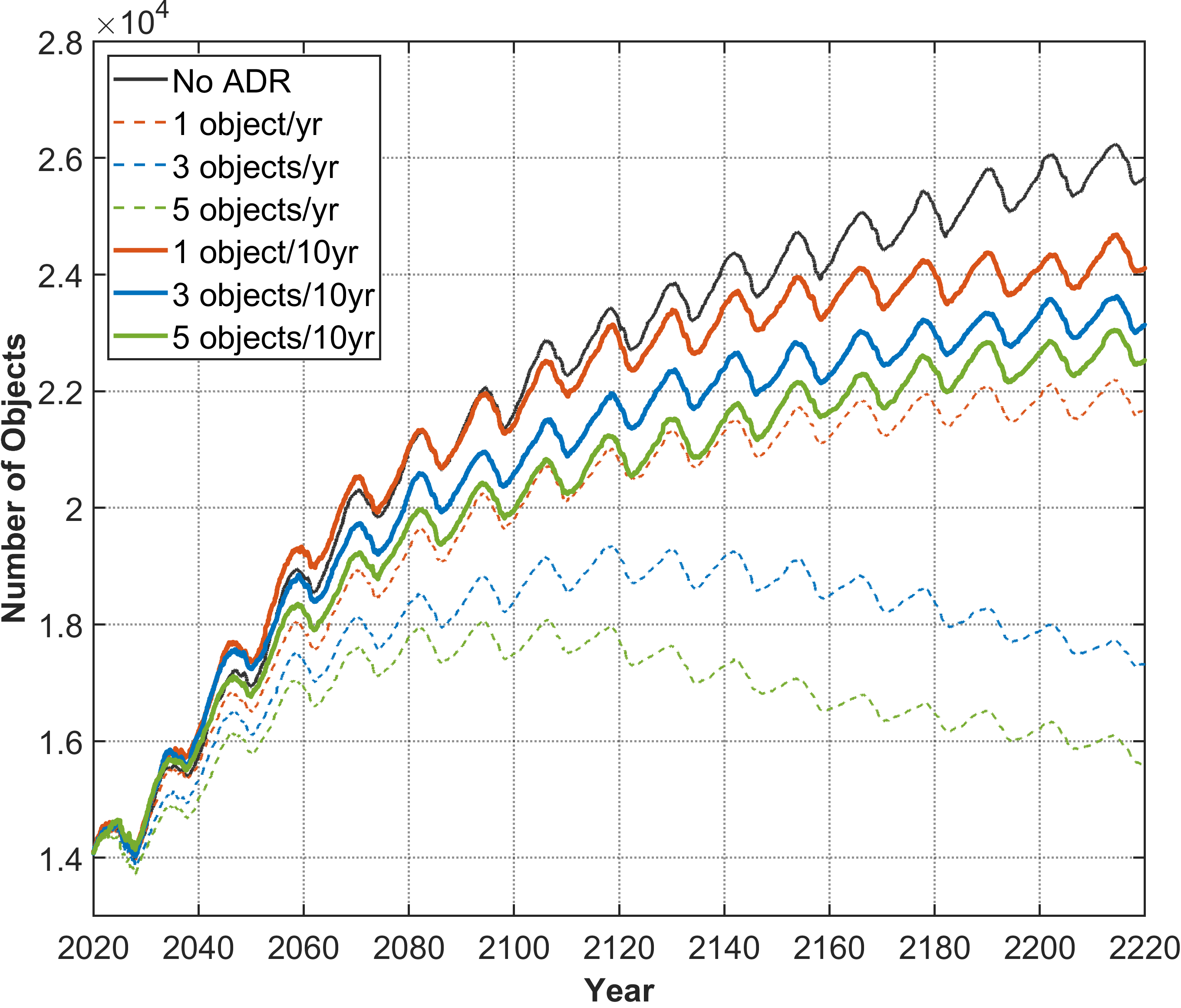}
        \label{fig:mitri_vs_10yr}
    }
    \caption{Effect of Varied Debris Removal Cadence on LEO Population Evolution using the MITRI index.}
    \label{fig:mitri_cadence_comparison}
\end{figure}
\vskip -1em
The rationale for this investigation stems from a fundamental trade-off. Longer intervals between removals may allow the risk profiles of objects to mature more fully, particularly the dynamic terms within the FMM index, potentially leading to a more optimal selection of truly long-term threats. Conversely, more frequent interventions, such as the annual cadence, allow the prompt removal of newly emerged high-risk objects before they can contribute to the debris population.

\begin{figure}[H]
    \centering
    \subfloat[Annual vs. 5-Year Cadence]{%
        \includegraphics[width=0.48\textwidth]{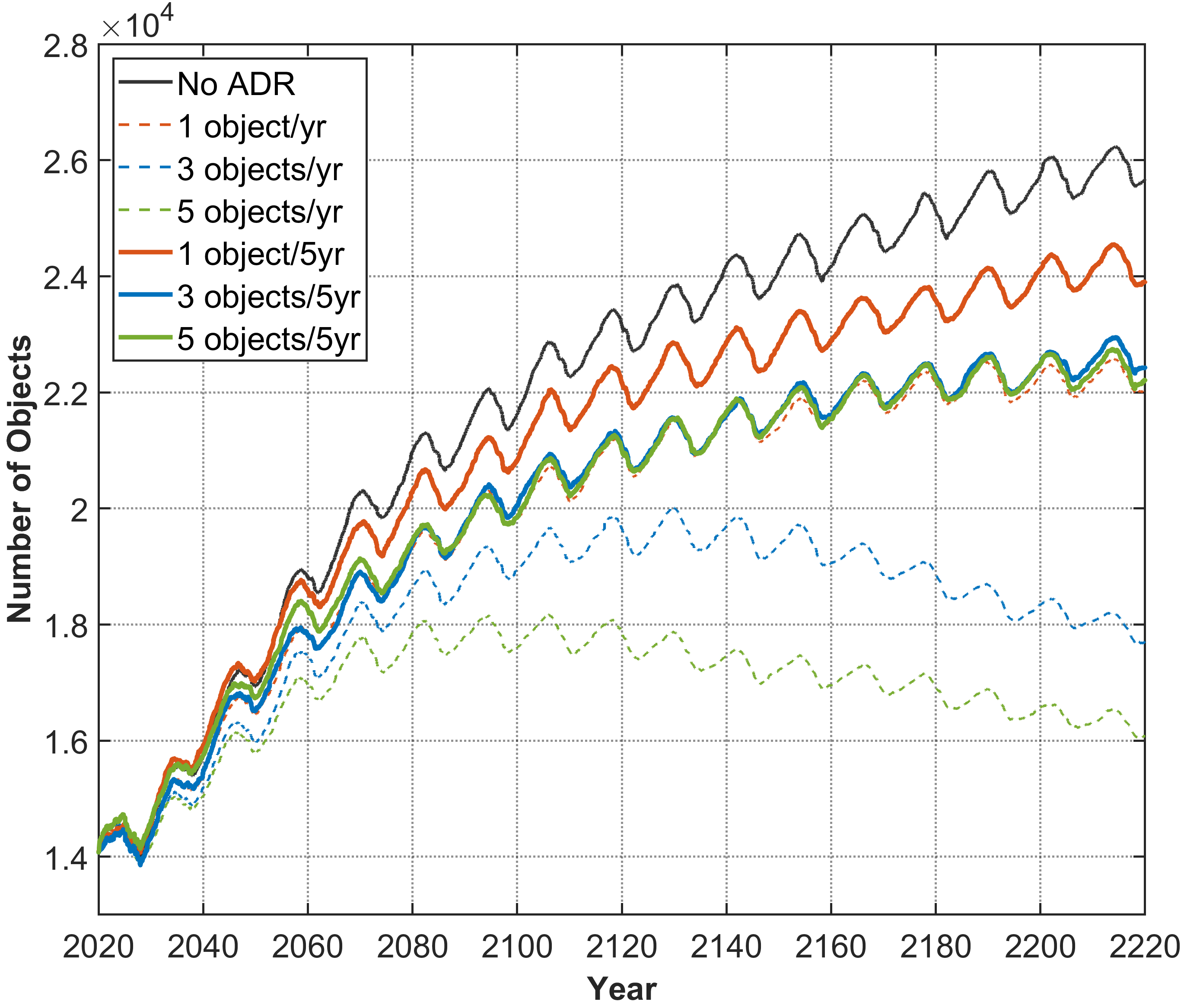}
        \label{fig:fmm_vs_5yr}
    }
    \hfill
    \subfloat[Annual vs. 10-Year Cadence]{%
        \includegraphics[width=0.48\textwidth]{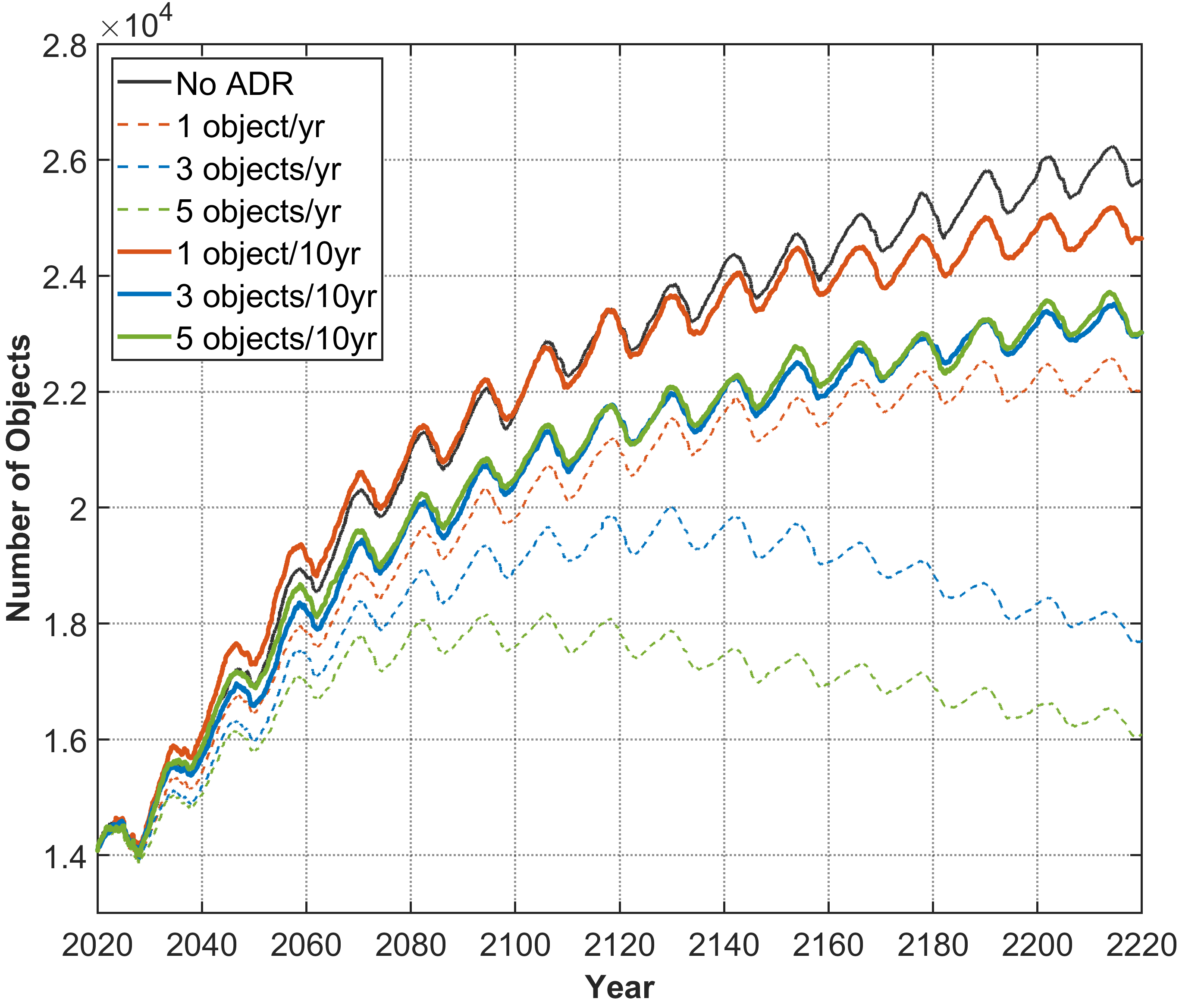}
        \label{fig:fmm_vs_10yr}
    }
    \caption{Effect of Varied Debris Removal Cadence on LEO Population Evolution using the FMM index.}
    \label{fig:fmm_cadence_comparison}
\end{figure}

The results for removals guided by the MITRI and FMM indices are presented in Figures~\ref{fig:mitri_cadence_comparison} and ~\ref{fig:fmm_cadence_comparison}, respectively. A clear and consistent trend is observable across all scenarios: a more frequent removal cadence is significantly more effective at stabilizing the debris environment. For any given removal rate $k$, the annual cadence consistently results in the lowest final object count, followed by the $5$-year cadence, which in turn outperforms the $10$-year cadence. This hierarchy holds for strategies guided by both indices, indicating that the strategic benefit of frequent intervention is a fundamental principle of effective ADR.
\vskip -1em
\vskip -1em
An interesting secondary finding emerges when comparing the performance of the two indices under these different cadences, revealing a nuanced trade-off. For the annual removal policy (solid lines), the MITRI index consistently leads to a lower final object count across all removal rates (comparing solid lines in Figure~\ref{fig:mitri_cadence_comparison} with those in Figure~\ref{fig:fmm_cadence_comparison}). Conversely, for the extended 5-year and 10-year cadences (dashed and dotted lines), this trend reverses, and the FMM index provides a marginally better outcome. This suggests that the optimal choice of risk index is context-dependent and should align with the operational timeline of the ADR campaign.

\subsubsection{Sensitivity Analysis of Mass-Related 
Parameters}\leavevmode\newline
The mass term, $(M/M_0)^{1.75}$, is a central component of the FMM index, as it is directly derived from empirical fragmentation models that establish a strong physical relationship between the mass of an object and its potential to generate catastrophic amounts of debris. To validate the criticality of this formulation and assess its sensitivity, two distinct experiments focused on mass-related parameters were conducted.

\begin{figure}[H]
    \centering
    \includegraphics[width=0.6\textwidth]{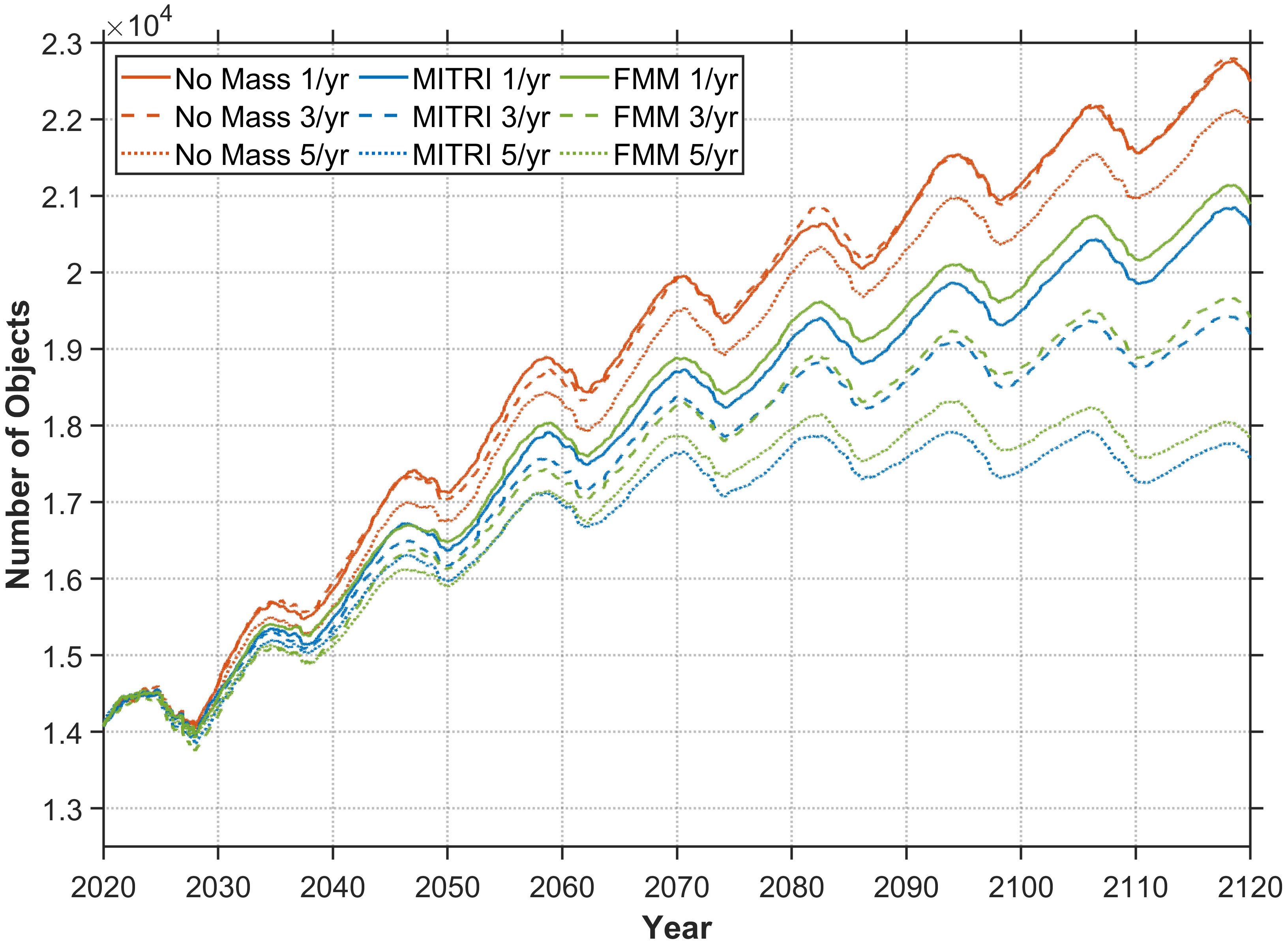}
    \caption{Impact of Removing the Mass Term on ADR Performance.}
    \label{fig:no_mass_comparison}
\end{figure}

The first experiment directly tested the importance of the mass term formulation by running two variant simulations over a 100-year period. The ``Linear Mass'' variant reduced the exponent from 1.75 to 1.0, modeling a simplified case where risk scales linearly with mass. The ``No Mass'' variant removed the term entirely, forcing the risk assessment to rely solely on orbital and probabilistic parameters.
The results unequivocally demonstrate that mass is an indispensable component for effective target prioritization. Figure~\ref{fig:no_mass_comparison} shows that completely removing the mass term leads to a catastrophic degradation in performance; the ``no mass'' strategy, even when removing 5 objects per year, results in a final population count significantly higher than even the baseline removal scenarios. This confirms that risk indices based solely on orbital parameters are insufficient. 
\vskip -1em
\vskip -1em

\begin{figure}[H]
    \centering
    \subfloat[MITRI index]{%
        \includegraphics[width=0.488\textwidth]{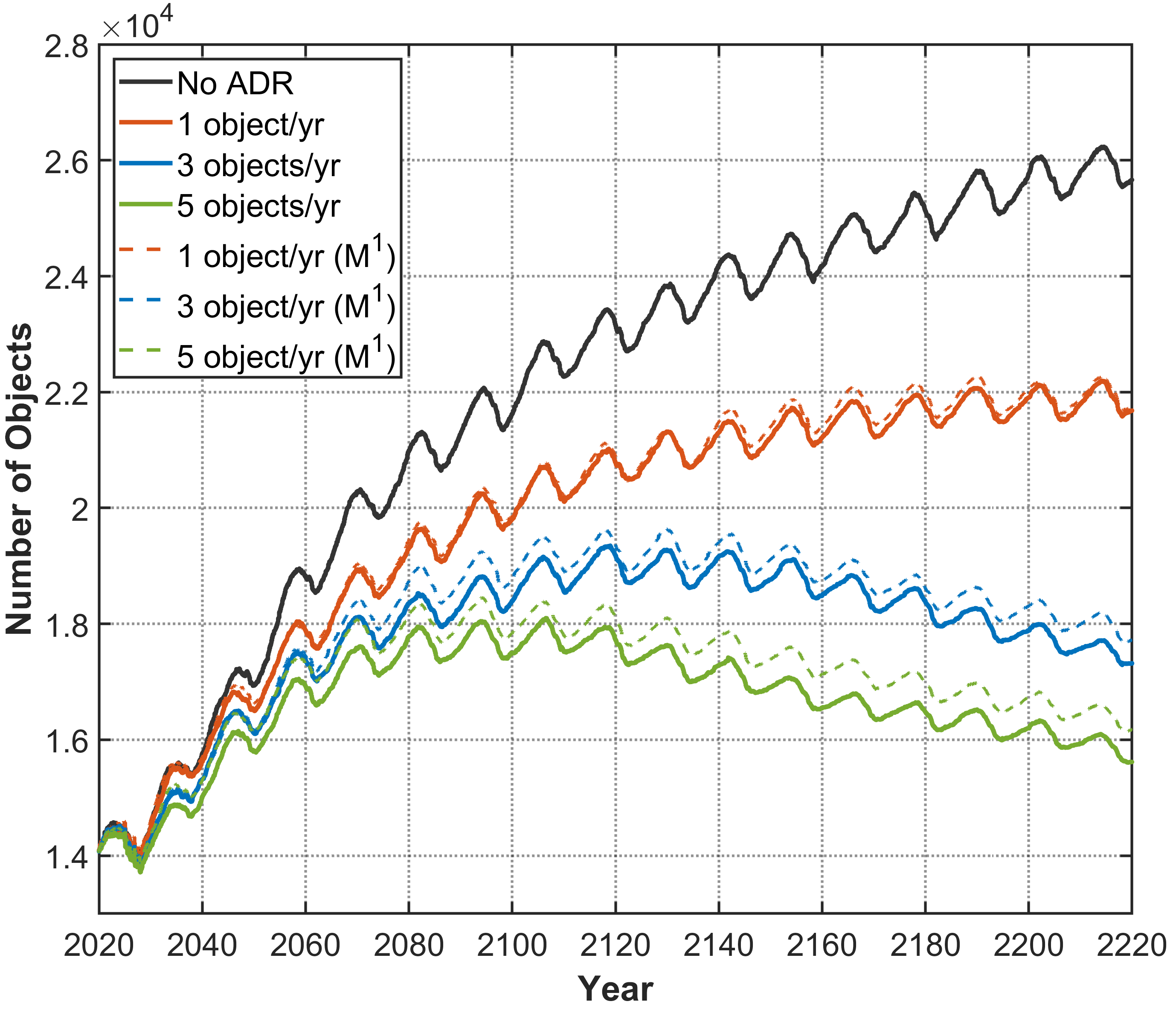}
        \label{fig:mitri_linear_mass}
    }
    \hfill
    \subfloat[FMM index]{%
        \includegraphics[width=0.488\textwidth]{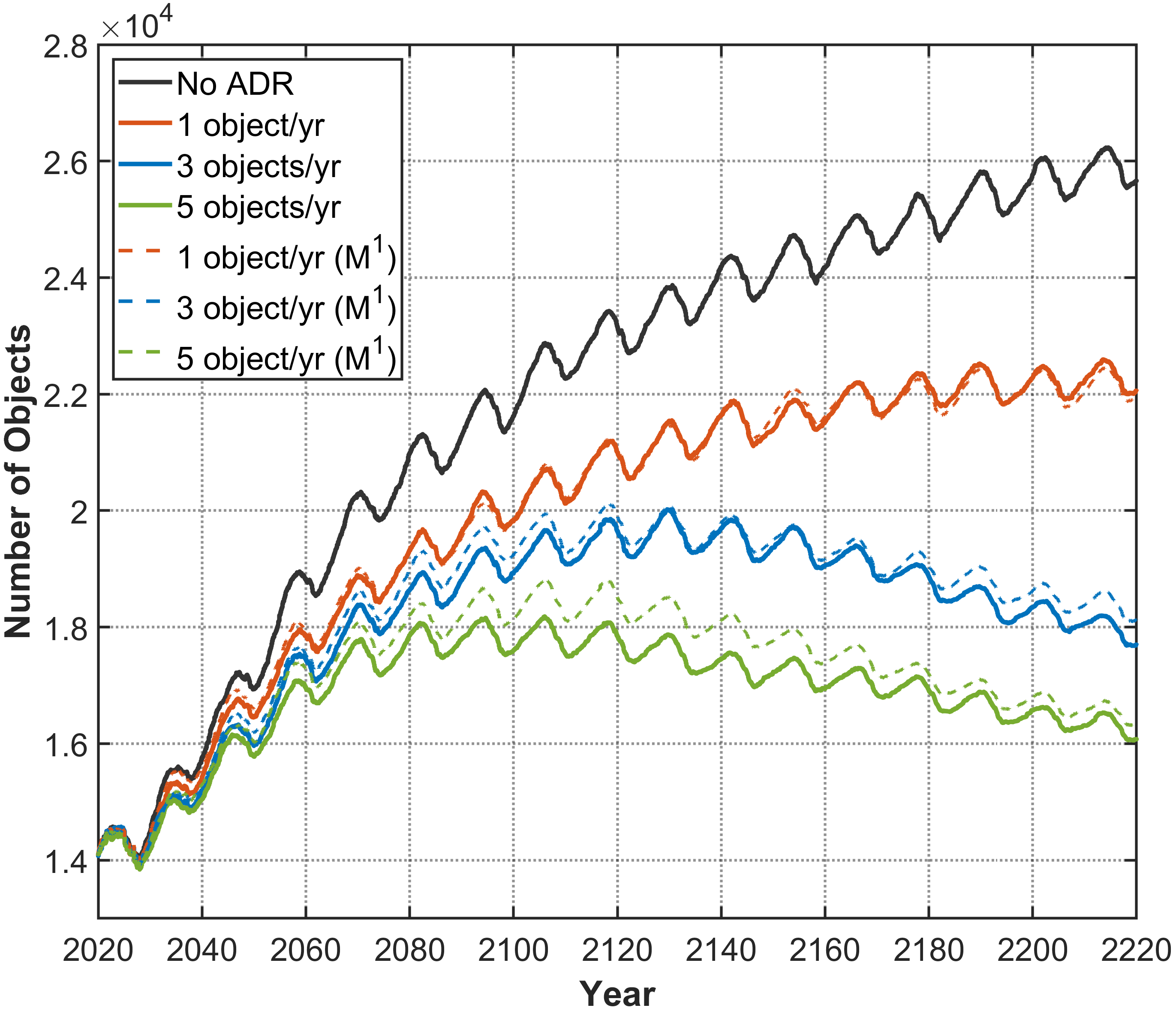}
        \label{fig:fmm_linear_mass}
    }
    \caption{Comparison of Baseline ($M^{1.75}$) vs. Linear Mass ($M^1$) Formulations.}
    \label{fig:linear_mass_comparison}
\end{figure}

Furthermore, Figure~\ref{fig:linear_mass_comparison} shows that using a linear mass term consistently underperforms relative to the baseline \(M^{1.75}\) formulation for both the MITRI and FMM indices. This validates that the 1.75 exponent, which accounts for both collision probability and the nonlinear fragmentation consequence, is a more effective and physically grounded formulation.
The second, complementary analysis examined the mass filtering threshold used within FMM's fictitious debris generation process. In this scenario, the mass filter was increased from the baseline threshold of \( > 10 \, \text{kg} \) to a more restrictive \( > 50 \, \text{kg} \). This experiment was designed to determine whether focusing ADR efforts exclusively on objects capable of generating the very largest fragments offers a more efficient risk mitigation strategy.
\vskip -1em
\vskip -1em
The results of this filtering experiment are shown in Figure~\ref{fig:filter_comparison}. The two approaches yield remarkably similar outcomes across all removal rates, with the final population counts for the 50 kg filter being only marginally higher than for the 10 kg baseline.
\subsubsection{Performance Metrics and Computational Cost}\leavevmode\newline
To ensure a consistent and objective evaluation across all numerical experiments, two primary performance metrics were defined to quantify the strategic effectiveness and practical efficiency of each ADR scenario.
\begin{enumerate}
    \item {Long-Term Population Stability:} The primary metric for strategic effectiveness is the total number of cataloged objects (\(>\)10 cm in size) remaining in LEO after the 200-year simulation period. This metric, visualized on the y-axis of Figures~\ref{fig:mitri_vs_rand} through~\ref{fig:filter_comparison}, directly reflects the overall collision risk and long-term orbital sustainability. A lower final object count indicates a more successful mitigation strategy.
\vskip -1em
\begin{figure}[H]
    \centering
    \includegraphics[width=0.6\textwidth]{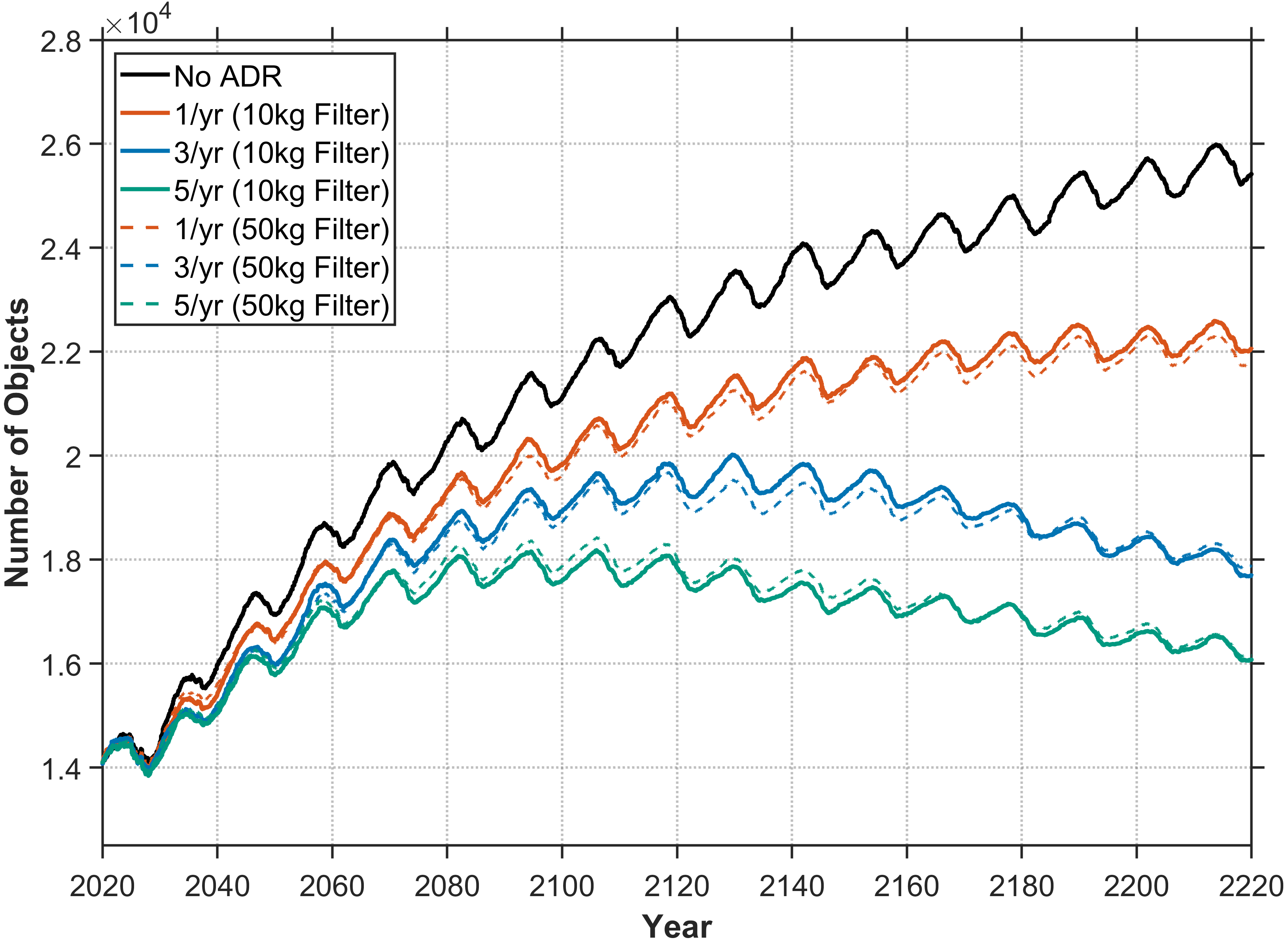}
    \caption{Comparison of FMM Performance with a 10~kg vs. 50~kg Debris Filter.}
    \label{fig:filter_comparison}
\end{figure}
\vskip -1em
    \item {Computational Cost:} The second metric evaluates the practical feasibility of each model variant by measuring its computational cost. This is quantified as the mean time required to complete a single Monte Carlo simulation run for a given scenario. This metric is crucial for assessing the trade-off between model complexity and performance, providing insight into the practicality of implementing these indices in operational settings.
\end{enumerate}

Although population stability is the key indicator of a model's effectiveness, its computational cost determines its practical utility. Table~\ref{tab:comp_cost} presents the results of the computational cost analysis. The most significant factor influencing cost is the choice between the MITRI and FMM frameworks. The data consistently show that FMM is more computationally intensive, with its static (1-time update) run being approximately ~1.54 times more expensive than the equivalent MITRI run. This increased demand is attributed directly to FMM's fictitious collision model, which processes every potential collision pair at each timestep. A secondary cost driver is the frequency of background density updates, which generally leads to a higher computational load for both indices. In contrast, most of the other sensitivity analyses had a negligible impact on computation time.

\vskip -1em
\begin{table}[H]
    \centering
    \caption{Mean Computational Cost per Simulation Seed.}
    \label{tab:comp_cost}
    \begin{tabular}{l l S[table-format=4.2] c}
        \toprule
        \textbf{Model / Scenario} & \textbf{Key Parameters} & {\textbf{Mean Time (s)}} & \textbf{Relative Cost} \\
        \midrule
        \multicolumn{4}{l}{\textit{Baseline Dynamic Indices}} \\
        MITRI & Static (1-time) & 1545.95 & 1.00x \\
              & 6-month update    & 1648.65 & 1.07x \\
              & 3-month update    & 1671.30 & 1.08x \\
              & 1-month update    & 2028.43 & 1.31x \\
        \addlinespace
        FMM   & Static (1-time) & 2384.26 & 1.54x \\
              & 6-month update    & 2580.99 & 1.67x \\
              & 3-month update    & 2361.35 & 1.53x \\
              & 1-month update    & 2485.14 & 1.61x \\
        \midrule
        \multicolumn{4}{l}{\textit{Sensitivity Analysis Variants}} \\
        FMM + Epsilon & $\varepsilon_1$ model    & 2629.05 & 1.70x \\
        FMM + Epsilon & $\varepsilon_2$ model    & 2423.93 & 1.57x \\
        Removal Cadence (MITRI) & 5-year interval & 1619.95 & 1.05x\\
        Removal Cadence (MITRI) & 10-year interval & 2000.53 & 1.29x\\
        Removal Cadence (FMM) & 5-year interval & 2622.28 & 1.69x\\
        Removal Cadence (FMM) & 10-year interval & 2386.86 & 1.54x\\
        No Mass Index & $M^0$ model     & 1388.93 & 0.90x \\
        MITRI (Linear Mass) & $M^1$ exponent    & 1664.95 &1.08x \\
        FMM (Linear Mass) & $M^1$ exponent    & 2437.96 & 1.58x \\
        FMM (\(\geq50\)kg Filter) & \(\geq50\) kg threshold & 2493.95 & 1.61x \\
        \bottomrule
    \end{tabular}
\end{table}
\vskip -1em

\section{DISCUSSION}\label{sec:discussion}
The analysis in Section 2 quantified the performance of different risk indices in various scenarios. This section now provides a higher-level interpretation of these findings, focusing on the specific performance trade-offs between the FMM and MITRI models and the structural importance of key parameters within the risk calculation.

\subsection{Interpretation of FMM Performance and Architectural Choices}\label{subsec:fmm_performance}
The experimental results demonstrated that FMM is a more precise tool for identifying high-risk targets for annual removal campaigns than its predecessors. This enhanced performance stems directly from its architectural shift to a proactive, potential-based risk calculation. By simulating debris generation for all high-probability conjunctions at every time step, the fictitious collision model captures the latent threat posed by objects in persistently dangerous orbits, a threat that stochastic, event-based models often miss until a collision is actually triggered.

Crucially, the sensitivity analyses confirm the robustness of this specific formulation. The attempt to introduce additional complexity via epsilon scaling, for instance, consistently degraded performance (increasing the final debris population by approximately 0.4 -- 7\% across all removal rates compared to the baseline FMM). A plausible explanation is that this scaling ``dilutes'' the risk index. 
By augmenting the Yearly Generated Debris ($D$) term for numerous low-probability conjunctions, the model overemphasizes general crowdedness at the expense of more critical, physically grounded parameters like mass ($M^{1.75}$) and residual lifetime ($L$). 
Consequently, the index is misled into prioritizing lower-consequence objects in congested orbits. This confirms that artificially inflating low-probability risks compromises overall mitigation efficacy.
Similarly, the analysis of the mass filtering threshold (Figure~\ref{fig:filter_comparison}) showed that the baseline 10~kg filter is well-calibrated, as the more restrictive 50~kg filter offered no significant performance benefit.

These findings strongly suggest that the FMM, in its proposed form, represents an optimized model. The data indicates that additional complexity is counterproductive, and the model is not overly sensitive to minor parameter changes, confirming its robustness for its intended mission.
\subsection{The Criticality of the Mass Term in Risk Assessment}\label{subsec:mass_discussion}

The results of the sensitivity analysis unequivocally demonstrate that mass is an indispensable component of an effective risk index. The catastrophic performance degradation observed when the mass term was removed entirely confirms that risk indices based solely on orbital or probabilistic parameters are insufficient for prioritizing high-consequence threats.

The physical justification for this is twofold. First, an object's mass serves as a direct proxy for its cross-sectional area, which linearly influences its probability of collision. 
Second, and more importantly, the fragmentation consequence of an impact is a direct function of the colliding masses, as empirically modeled by the NASA SBM. The $M^{1.75}$ formulation in the baseline indices correctly captures both of these effects. 
The observed underperformance of the $M^1$ (linear mass) variant further validates this, proving that accounting for the nonlinear fragmentation consequence ($M^{0.75}$) is critical for accurate risk prediction. 
Therefore, the data confirm a fundamental principle for ADR strategy: any robust prioritization must be fundamentally anchored by a physically grounded mass term. Models that neglect or oversimplify this component will fail to accurately predict the most significant long-term threats.

\subsection{Implications of Removal Cadence on Index Selection}\label{subsec:cadence_discussion}

A significant finding from this study is the nuanced performance trade-off between the MITRI and FMM indices that emerges when the removal cadence is varied. For frequent, annual removals, MITRI proved to be the more effective model, while for less frequent, extended cadences of 5 and 10 years, FMM demonstrated a marginal performance advantage.

This trade-off highlights a fundamental difference in their risk philosophies. MITRI, being event-based, excels as a tactical tool. It is designed to react to specific high-probability collisions detected in near-real time, making it highly effective at identifying the most immediate, near-term threat for an annual removal campaign.

In contrast, FMM's fictitious collision model operates as a strategic tool. It continuously aggregates potential risk from all nearby objects over extended periods, creating a more stable, time-averaged hazard assessment. For long-range strategic planning with infrequent removals (e.g., $5-10$ year cadences), this integrated risk assessment provides a marginally better prediction of long-term consequences, as evidenced by the slightly lower debris populations in those scenarios. 
The practical implication for mission architects is that the optimal index is context-dependent. For missions involving frequent, annual interventions, MITRI's ability to pinpoint immediate threats is superior. For long-term strategic campaigns with extended intervals between removals, the cumulative approach of FMM offers a slight advantage in predicting future risk evolution.
\subsection{Computational Cost-Benefit Analysis}\label{subsec:cost_benefit}
The higher computational cost of the FMM framework, established in the previous analysis, requires a clear cost-benefit evaluation. The crucial question is whether this computational overhead is justified by a proportional increase in performance.

For the primary use case of strategically planning annual ADR campaigns, the answer is yes. The near-perfect identification rate of FMM (98–100\%) for high-risk targets represents a significant improvement over MITRI. Given that real-world ADR missions require months or even years of preparation, the additional simulation time is a negligible price for substantially higher confidence in target selection.

However, this conclusion is context-dependent. For time-critical applications where low latency is important, such as a rapid response to a recent breakup event, MITRI's faster, event-based architecture would remain the more suitable tool. This reinforces the finding that the choice between the two indices is a deliberate one that should be aligned with the specific operational needs and timelines of the mission.

\section{Conclusion}\label{sec:conclusion}
This study presented the development and comprehensive evaluation of an enhanced risk index, FMM, for prioritizing targets for active debris removal. 
Using the MOCAT-MC simulation framework, we conducted a series of numerical experiments to validate the index against existing models and probe its formulation's robustness through a wide-ranging sensitivity analysis.

The results confirm that a targeted, risk-based ADR strategy is fundamentally more effective than random removal at ensuring the long-term stability of the LEO environment. 
The key findings of this work demonstrate that the FMM index, which incorporates a fictitious collision model and a dynamic background density within a concurrent analysis framework, has a superior ability to identify high-risk objects compared to its predecessors. This enhanced performance is anchored by the inclusion of a physically grounded mass term, \(M^{1.75}\), which the sensitivity analysis proved to be an indispensable component of any effective risk calculation. Furthermore, the analysis revealed a critical operational trade-off, leading to the practical recommendation that mission architects should select indices contextually: MITRI is more effective for frequent annual removals, while FMM has a marginal advantage for less frequent long-term strategic campaigns.

This study also lays the foundation for a more thorough risk assessment in subsequent research. The existing framework effectively identifies high-risk parent objects; however, a later post-processing study could evaluate the long-term threat posed by their fragmentation products, or ``children." This investigation would involve propagating the debris cloud created by a potential collision involving a high-ranking object to analyze the long-term evolution of the resultant fragments. The initial ranking could be improved by considering the overall environmental effect of this secondary debris population. This would allow for an even more optimal prioritization of ADR targets whose removal would mitigate not only the immediate threat posed by the parent object but also the long-term cascading threat from its potential debris.

Building on the potential for a post-analysis of fragmentation products, future work should also focus on two other key areas. First, a critical next step is to validate the FMM's rankings against on-orbit targets from planned ADR missions. Second, machine learning techniques could be explored to develop a computationally efficient surrogate model of FMM, aiming to reduce its runtime while preserving its high-fidelity risk assessment.

\section{Acknowledgement}\label{sec:acknowledgment}
This publication is based on work supported by the NASA-ROSES program through grant number\linebreak~80NSSC24K1768.

\bibliographystyle{AAS_publication}
\bibliography{references}

\end{document}